\documentclass[showpacs,prd]{revtex4}
\usepackage{pifont}
\usepackage{bbding}
\usepackage{}
\usepackage{mathrsfs}
\usepackage{longtable,lscape} \usepackage{multirow}
\usepackage{txfonts,bm}
\usepackage{amssymb}
\usepackage{indentfirst}
\usepackage{graphicx,booktabs}
\usepackage{color}
\usepackage{hyperref}

\def\n{\nonumber}
\def\f{\frac}

\def\L{\Lambda}

\begin{document}

\title{Isospin breaking, coupled-channel effects, and X(3872)}
\author{Ning Li}\email{leening@pku.edu.cn}
\affiliation{Department of Physics
and State Key Laboratory of Nuclear Physics and Technology\\
Peking University, Beijing 100871, China}
\author{Shi-Lin Zhu}
\email{zhusl@pku.edu.cn} \affiliation{Department of Physics
and State Key Laboratory of Nuclear Physics and Technology\\
and Center of High Energy Physics, Peking University, Beijing
100871, China }
\date{\today}

\begin{abstract}

We re-investigate the possibility of $X(3872)$ as a $D\bar{D}^*$
molecule with $J^{PC}=1^{++}$ within the framework of both the
one-pion-exchange (OPE) model and the one-boson-exchange (OBE)
model. After careful treatment of the S-D wave mixing, the mass
difference between the neutral and charged $D(D^*)$ mesons and the
coupling of the $D\bar{D}^*$ pair to $D^*\bar{D}^*$, a loosely bound
molecular state $X(3872)$ emerges quite naturally with large isospin
violation in its flavor wave function. For example, the isovector
component is $26.24\%$ if the binding energy is 0.30 MeV, where
the isospin breaking effect is amplified by the tiny binding
energy. After taking into account the phase space difference and
assuming the $3\pi$ and $2\pi$ come from a virtual omega and rho
meson respectively, we obtain the ratio of these two hidden-charm
decay modes: $\mathcal{B}(X(3872)\rightarrow \pi^+\pi^-\pi^0
J/\psi)/\mathcal{B}(X(3872)\rightarrow \pi^+\pi^- J/\psi)=0.42$
for the binding energy being 0.3 MeV, which is consistent with
the experimental value.

\end{abstract}


\maketitle

\section{Introduction} \label{Introduction}

In 2003, the Belle Collaboration observed a narrow charmonium-like
state $X(3872)$ in the exclusive decay process $B^{\pm}\rightarrow
XK^{\pm}$ followed by $X\rightarrow
\pi^+\pi^-J/\psi$~\cite{Abe:2003hq}. Later, this state was
confirmed by CDF~\cite{Acosta:2003zx}, D0~\cite{Abazov:2004kp} and
BABAR~\cite{Aubert:2004fc}. The current value of the $X(3872)$
mass is $M_{X(3872)}=(3871.95\pm0.48(\mbox{stat})\pm0.12(\mbox{syst}))$
MeV~\cite{Aaij:2011sn} while the updated value of the width of
$X(3872)$ is $\Gamma_{X(3872)}<1.2$ MeV~\cite{Choi:2011fc}. Due to
its exotic properties, the $X(3872)$ state has attracted much
attention since its
discovery
~\cite{Braaten:2004jg,Matheus:2006xi,Chiu:2006hd,Liu:2008qb,Gamermann:2009fv,Matheus:2009vq,DeSanctis:2011zza,Valderrama:2012jv,Yang:2012my}.
Despite huge efforts, the nature of
$X(3872)$ is still unclear. Up to now, the proposed
interpretations of the $X(3872)$ include ``hadronic
molecule"~\cite{Swanson:2003tb,Tornqvist:2004qy,AlFiky:2005jd,Thomas:2008ja,Liu:2008tn,Lee:2009hy},
$c\bar{c}g$ hybrid~\cite{Li:2004sta},
tetraquark~\cite{Maiani:2004vq}
charmonium~\cite{Barnes:2003vb,Suzuki:2005ha}.

Experimentally, both BABAR~\cite{Aubert:2005vi} and
BELLE~\cite{Choi:2011fc} did not find the charged partner of the
$X(3872)$, which suggests that the $X(3872)$ is an isoscalar.
However, the Belle Collaboration reported the branching fraction
ratio, $\mathcal{B}(X\to\pi^+\pi^-\pi^0
J/\psi)/\mathcal{B}(X\to\pi^+\pi^-J/\psi)
  =1.0\pm0.4(\mbox{stat})\pm0.3(\mbox{syst})$~\cite{Abe:2005ix}, which indicates that there
exists large isospin breaking for the hidden-charm decay of
$X(3872)$. This result was later confirmed by the BABAR
collaboration, $\mathcal{B}(X\to\pi^+\pi^-\pi^0
J/\psi)/\mathcal{B}(X\to\pi^+\pi^-J/\psi)=0.8\pm0.3$~\cite{delAmoSanchez:2010jr}.
In addition, the charge parity of $X(3872)$ is even ($C=+1$) from
its radiative decay $X(3872)\to\gamma
J/\psi$~\cite{Abe:2005ix,Aubert:2006aj}. The quantum numbers of
$X(3872)$ are probably $J^{PC}=1^{++}$ or
$2^{-+}$~\cite{Abe:2005iya,Abulencia:2006ma}.

The proximity of the $X(3872)$ to the $D^0\bar{D}^{*0}$ threshold
strongly suggests that the $X(3872)$ might be a weakly bound
$D^0\bar{D}^{*0}$ molecule. If the $X(3872)$ is really a loosely
bound $D^0\bar{D}^{*0}$ molecule, we expect that the long-range
pion exchange plays a dominant role among the exchanged mesons
since the constituent hadrons of the hadronic molecule should be
well-separated. We also expect that there exists strong mixing
between $D^0\bar{D}^{*0}$ and $D^+D^{*-}$ due to the closeness of
the threshold of $D^0\bar{D}^{*0}$ and $D^+D^{*-}$. Actually, if
one only considers the neutral $D\bar{D}^*$ pair, the interaction
strength is only one third of that of the isospin singlet. On the
other hand, compared with the small binding energy of $X(3872)$
(less than 1 MeV), the mass difference between $D^0\bar{D}^{*0}$
and $D^+D^{*+}$ ($\sim8.1$ MeV) is so large that the large isospin
breaking may occur for the $X(3872)$. Further more, the coupling
of $D\bar{D}^*$ to $D^*\bar{D}^*$ should also affect the binding
of $X(3872)$ since the mass difference is about
$m_{D^*\bar{D}^*}-m_{X(3872)}\simeq 140$ MeV only.

In the present paper, we shall take into account the S-D wave
mixing which plays an important role in forming the loosely bound
deuteron, the charged $D\bar{D}^*$ pair, the mass difference
between the neutral and the charged $D(\bar{D}^*)$ meson and the
coupling of $D\bar{D}^*$ to $D^*\bar{D}^*$. In order to highlight
the contribution of the long-range pion exchange, we first study
the system with the pion exchange alone. Then we move on and
include the other light meson exchanges with the OBE framework.
Since the quantum numbers of $X(3872)$ have not been determined
exactly, we investigate both the $J^{PC}=1^{++}$ and $2^{-+}$ cases
within the ``hadronic molecule" framework.

The paper is organized as follows. After the introduction
\ref{Introduction}, we present the formalism including the
lagrangians and the effective potentials in Section~\ref{X3872}
and the numerical results in Section~\ref{numerical}. We discuss
the isospin symmetry breaking in Section~\ref{isospin}. We
summarize our results and conclusions in Section~\ref{Summary}. We
list some useful formulae and the discussion of the possible
$J^{PC}=1^{-+}$ molecular state in the APPENDIX.

\section{$X(3872)$ as a Hadronic Molecule}\label{X3872}

The proximity of the $X(3872)$ to the threshold of
$D^0\bar{D}^{*0}$ strongly suggests that the $X(3872)$ is probably
a loosely bound $D^0\bar{D}^{*0}$ molecule. In the present work,
we investigate this probability. Given the $J^{PC}$ assignment of
the $X(3872)$ has not been exactly measured experimentally, we
consider both the $1^{++}$ and $2^{-+}$ cases. However, as we will
show below, we do not find a binding solution for the $2^{-+}$
case with cutoff parameter less than 2.0 GeV.

We want to find out the specific role of the charged $D\bar{D}^*$
mode, the isospin breaking and the coupling of the $X(3872)$ to
$D^*\bar{D}^*$ in forming the loosely bound $X(3872)$. We first
consider the neutral component $D^0\bar{D}^{*0}$ only and include
the S-D wave mixing, which corresponds to Case I. Then we add
the charged $D^+D^{*-}$ component to form the exact $D\bar{D}^*$
isospin singlet with the S-D mixing, which is Case II. Since the
$1^{++}$ $D^*\bar{D}^*$ channel lies only 140 MeV above and
couples strongly to the $D\bar{D}^*$ channel, we further introduce
the coupling of $D\bar{D}^*$ to $D^*\bar{D}^*$ in Case III.
Finally, we move one step further and take into account the
explicit mass splitting between the charged and neutral
$D(D^\ast)$ mesons, which is the physical Case IV. We list the
channels of these four cases in Table~\ref{Channel}.

In Case IV we consider the isospin breaking for $D\bar{D}^*$
only but keep the isospin limit for the $D^*\bar{D}^*$ channel.
Since the threshold of $D^*\bar{D}^*$ is about 140 MeV above the
$X(3872)$ mass, the probability of the $D^*\bar{D}^*$ component is
already quite small due to such a large mass gap. The isospin
breaking effect due to the mass splitting of the $D^*\bar{D}^*$
pair is even smaller and negligible. In Case IV we have omitted
the channel ${1\over
\sqrt{2}}\left(D^{*0}\bar{D}^{*0}+D^{*+}D^{*-}\right)|^5D_1>$. At
the first glimpse, this channel should also be included. After
careful calculation, it turns out that the matrix elements between
this channel and other channels are zero.

\begin{table}
\renewcommand{\arraystretch}{1.0} \caption{The different channels for Cases
I, II, III and IV of $X(3872)$ with $J^{PC}=1^{++}$. For
simplicity, we adopt the following short-hand notations,
$\left[D^{0}\bar{D}^{*0}\right]\equiv {1\over
\sqrt{2}}\left(D^{0}\bar{D}^{*0}-D^{*0}\bar{D}^{0}\right)$,
$\left[D^+D^{*-}\right]\equiv{1\over
\sqrt{2}}\left(D^+D^{*-}-D^{*+}D^-\right)$,
$\left\{D^*\bar{D}^*\right\}\equiv {1\over
\sqrt{2}}\left(D^{*0}\bar{D}^{*0}+D^{*+}D^{*-}\right)$ and
$\left(D\bar{D}^*\right) \equiv {1\over
2}\left[\left(D^{0}\bar{D}^{*0}-D^{*0}\bar{D}^{0}\right)
+\left(D^{+}D^{*-}-D^{*+}D^{-}\right)\right]$. ``$ -$ " means the
corresponding channel does not exist.}\label{Channel}
\begin{tabular*}{18cm}{@{\extracolsep{\fill}}ccccccc}
\toprule[0.8pt] \toprule[0.8pt] \addlinespace[2pt]
      ~       & \multicolumn{6}{c}{Channels}\\
      Cases   &   1   &   2   &   3   &    4   &   5   &   6  \\
\specialrule{0.6pt}{1pt}{3pt}
       I      &$\left[D^{0}\bar{D}^{*0}\right]|^3S_1>$&
              $\left[D^{0}\bar{D}^{*0}\right]|^3D_1>$ &
               $ - $  & $-$   &  $-$  &  $-$  \\ [3pt]
       II     &$\left(D\bar{D}^{*}\right)|^3S_1>$&
               $\left(D\bar{D}^{*}\right)|^3D_1>$&
               $ - $  & $-$   &  $-$  &  $-$  \\ [3pt]
       III    & $\left(D\bar{D}^{*}\right)|^3S_1>$&
                $\left(D\bar{D}^{*}\right)|^3D_1>$ &
                $-$& $-$&
               $\left\{D^*\bar{D}^*\right\}|^3S_1>$&
               $\left\{D^*\bar{D}^*\right\}|^3D_1>$\\ [3pt]
       IV(Phy)&$\left[D^0\bar{D}^{*0}\right]|^3S_1>$&
              $\left[D^0\bar{D}^{*0}\right]|^3D_1>$ &
              $\left[D^+D^{*-}\right]|^3S_1>$&
              $\left[D^+D^{*-}\right]|^3D_1>$&
              $\left\{D^*\bar{D}^*\right\}|^3S_1>$&  $\left\{D^*\bar{D}^*\right\}|^3D_1>$\\ [3pt]
\bottomrule[0.8pt] \bottomrule[0.8pt]
\end{tabular*}
\end{table}

\subsection{The Lagrangians and The Coupling Constants}\label{Lagrangian}

The lagrangians with the heavy quark symmetry and the chiral
symmetry read
~\cite{Falk:1992cx,PhysRevD.45.R2188,Yan1992,Grinstein:1992qt,Cheng:1992xi,Casalbuoni:1996pg,Sun2011,Ding2009a}

\begin{eqnarray}
\mathcal{L}_{P^{(*)}P^{(*)}M} &=&
-i\frac{2g}{f_\pi}\varepsilon_{\alpha\mu\nu\lambda}
v^\alpha P^{*\mu}_{b}\partial^\nu M_{ba}{P}^{*\lambda\dag}_{a}
+i \frac{2g}{f_\pi}\varepsilon_{\alpha\mu\nu\lambda}
v^\alpha\widetilde{P}^{*\mu\dag}_{a}\partial^\nu{}M_{ab}\widetilde{P}^{*\lambda}_{b}
\nonumber\\
&~&-\frac{2g}{f_\pi}(P_bP^{*\dag}_{a\lambda}+
P^{*}_{b\lambda}P^{\dag}_{a})\partial^\lambda{}
M_{ba}
+\frac{2g}{f_\pi}(\widetilde{P}^{*\dag}_{a\lambda}\widetilde{P}_b+
\widetilde{P}^{\dag}_{a}\widetilde{P}^{*}_{b\lambda})\partial^\lambda M_{ab}.\label{Lagrangian:p}
\end{eqnarray}

\begin{eqnarray}
  \mathcal{L}_{P^{(*)}P^{(*)}V}
  &=& -\sqrt{2}\beta{}g_VP_b v\cdot\hat{\rho}_{ba}P_a^{\dag}
  +\sqrt{2}\beta g_V\widetilde{P}^{\dag}_a v\cdot\hat{\rho}_{ab}
  \widetilde{P}_b \nonumber\\
  &~&- 2\sqrt{2}\lambda{}g_V \varepsilon_{\lambda\mu\alpha\beta}v^\lambda
  (P^{}_bP^{*\mu\dag}_a +
  P_b^{*\mu}P^{\dag}_a)
  (\partial^\alpha{}\hat{\rho}^\beta)_{ba}
 -2\sqrt{2}\lambda{}g_V
\varepsilon_{\lambda\mu\alpha\beta}v^\lambda
(\widetilde{P}^{*\mu\dag}_a\widetilde{P}^{}_b
+\widetilde{P}^{\dag}_a\widetilde{P}_b^{*\mu})
  (\partial^\alpha{}\hat{\rho}^\beta)_{ab}\nonumber\\
  &~&+\sqrt{2}\beta{}g_V P_b^{*}\cdot P^{*\dag}_a
  v\cdot\hat{\rho}_{ba}
  -i2\sqrt{2}\lambda{}g_V P^{*\mu}_b (\partial_\mu{}
  \hat{\rho}_\nu - \partial_\nu{}\hat{\rho}_\mu)_{ba}P^{*\nu\dag}_a
  \nonumber\\
  &~&-\sqrt{2}\beta g_V
  \widetilde{P}^{*\dag}_a\cdot\widetilde{P}_b^{*}
  v\cdot\hat{\rho}_{ab}-i2\sqrt{2}\lambda{}g_V\widetilde{P}^{*\mu\dag}_a(\partial_\mu{}
  \hat{\rho}_\nu - \partial_\nu{}\hat{\rho}_\mu)_{ab}\widetilde{P}^{*\nu}_b.\label{lagrangian:v}
  \end{eqnarray}

  \begin{eqnarray}
  \mathcal{L}_{P^{(*)}P^{^(*)}\sigma}
  &=& -2g_s P^{}_b\sigma P^{\dag}_b
 -2g_s\widetilde{P}^{\dag}_b\sigma\widetilde{P}_b \nonumber \\
 &~&+2g_sP^{*}_b\cdot{}P^{*\dag}_b\sigma
 +2g_s\widetilde{P}^{*\dag}_b\cdot{}\widetilde{P}^{*}_b\sigma.\label{lagrangian:s}
\end{eqnarray}
$P=\left(D^0,D^+,D_s^+\right)$ and $P^*=\left(D^{*0},D^{*+},D_s^{*+}\right)$ are the
heavy meson fields
while $\widetilde{P}=\left(\bar{D}^0,D^-,D_s^-\right)$ and
$\widetilde{P}^*=\left(\bar{D}^{*0},D^{*-},D_s^{*-}\right)$ are the
heavy anti-meson fields.
The exchanged pseudoscalar meson and vector meson matrices $M$ and
$\hat{\rho}^{\mu}$ are defined as
\begin{eqnarray}
  M=\left(\begin{array}{ccc}
{\pi^0 \over \sqrt{2}}+{\eta \over \sqrt{6}} & \pi^+      &     K^+\\
\pi^-  & -{\pi^0\over \sqrt{2}}+{\eta \over \sqrt{6}} & K^0  \\
K^-    &\bar{K}^0 & -{2\over \sqrt{6}} \eta \\
\end{array}
  \right),\quad
  \hat{\rho}^{\mu}=\left(\begin{array}{ccc}
  {\rho^0 \over \sqrt{2}}+{\omega \over \sqrt{2}} & \rho^+      &     K^{*+}\\
\rho^-  & -{\rho^0\over \sqrt{2}}+{\omega \over \sqrt{2}} & K^{*0}  \\
K^{*-}  &\bar{K}^{*0} & \phi \\
\end{array}\right)^{\mu}.
\end{eqnarray}

In the OPE model,
 there are two coupling constants $f_{\pi}$ and $g$. $f_{\pi}=132$ MeV is the pion
 decay constant. The coupling constants $g$ was studied by many theoretical approaches,
 such as quark model~\cite{Falk:1992cx} and QCD sum rule~\cite{Dai:1998vh,Navarra:2000ji}.
Here, we take the experimental result of the CLEO Collaboration, $g=0.59\pm0.07\pm0.01$,
which was extracted from the full width of $D^{*+}$
~\cite{Ahmed:2001xc}. Following~\cite{Isola:2003fh,Bando:1987br}, the
parameters related to the vector meson exchange are $g_v=5.8$ and
$\beta=0.9$ determined by the vector meson dominance mechanism,
and $\lambda=0.56~\mbox{GeV}^{-1}$ obtained by matching the form factor
predicted by the effective theory approach with that obtained by the light cone sum
rule and the lattice QCD.
The coupling constant for the scalar meson exchange is
$g_s=g_{\pi}/(2\sqrt{6})$~\cite{Liu2008a} with $g_{\pi}=3.73$.
We summarize the parameters used in our calculation in Table~\ref{Parameter}.

\begin{table}[htp]
\renewcommand{\arraystretch}{1.0} \caption{The coupling constants
and the masses of the heavy mesons and the exchanged light mesons used in our calculation. The masses of
the mesons are taken from the PDG~\cite{pdg2010}. For the channel
$\left\{D^*\bar{D}^*\right\}$, we keep the isospin symmetry and adopt
$m_{D^*}=(m_{D^{*\pm}}+m_{D^{*0}})/2=2008.6$ MeV and
$m_{\pi}=(m_{\pi^{\pm}}+m_{\pi^0})/2=137.27$ MeV.}\label{Parameter}
\begin{tabular*}{18cm}{@{\extracolsep{\fill}}ccc|cc}
\toprule[0.8pt] \toprule[0.8pt] \addlinespace[2pt]
 \multicolumn{3}{c|}{Coupling Constants} & \multicolumn{2}{c}{Masses (MeV)}\\
 Pseudoscalar  &  Vector      &    Scalar      &  Heavy Mesons  & Exchanged Mesons  \\
\specialrule{0.6pt}{1pt}{3pt}
$g=0.59$       &  $g_v=5.8$   &$g_s={g_{\pi}\over 2\sqrt{6}}$ with $g_{\pi}=3.73$ &
$m_{D^{\pm}}=1869.60$  &  $m_{\pi^{\pm}}=139.57$ \\
$f_{\pi}=132$ MeV      & $\beta=0.9$   &   ~      &
$m_{D^0}=1864.83$      & $m_{\pi^0}=134.98$      \\
        ~              & $\lambda=0.56~\mbox{GeV}^{-1}$&       ~                  &
 $m_{D^{*\pm}}=2010.25$& $m_{\eta}=547.85$       \\
        ~              &           ~             &             ~                  &
 $m_{D^{*0}}=2006.96$  & $m_{\rho}=775.49$       \\
        ~              &           ~             &             ~                  &
        ~              & $m_{\omega}=782.65$     \\
        ~              &           ~             &              ~                 &
        ~              & $m_{\sigma}=600$        \\
\bottomrule[0.8pt] \bottomrule[0.8pt]
\end{tabular*}
\end{table}

\subsection{The Effective Potentials}\label{Effect:potential}

\begin{figure}[htp]
\centering
\begin{tabular}{cc}
\includegraphics[width=0.45\textwidth]{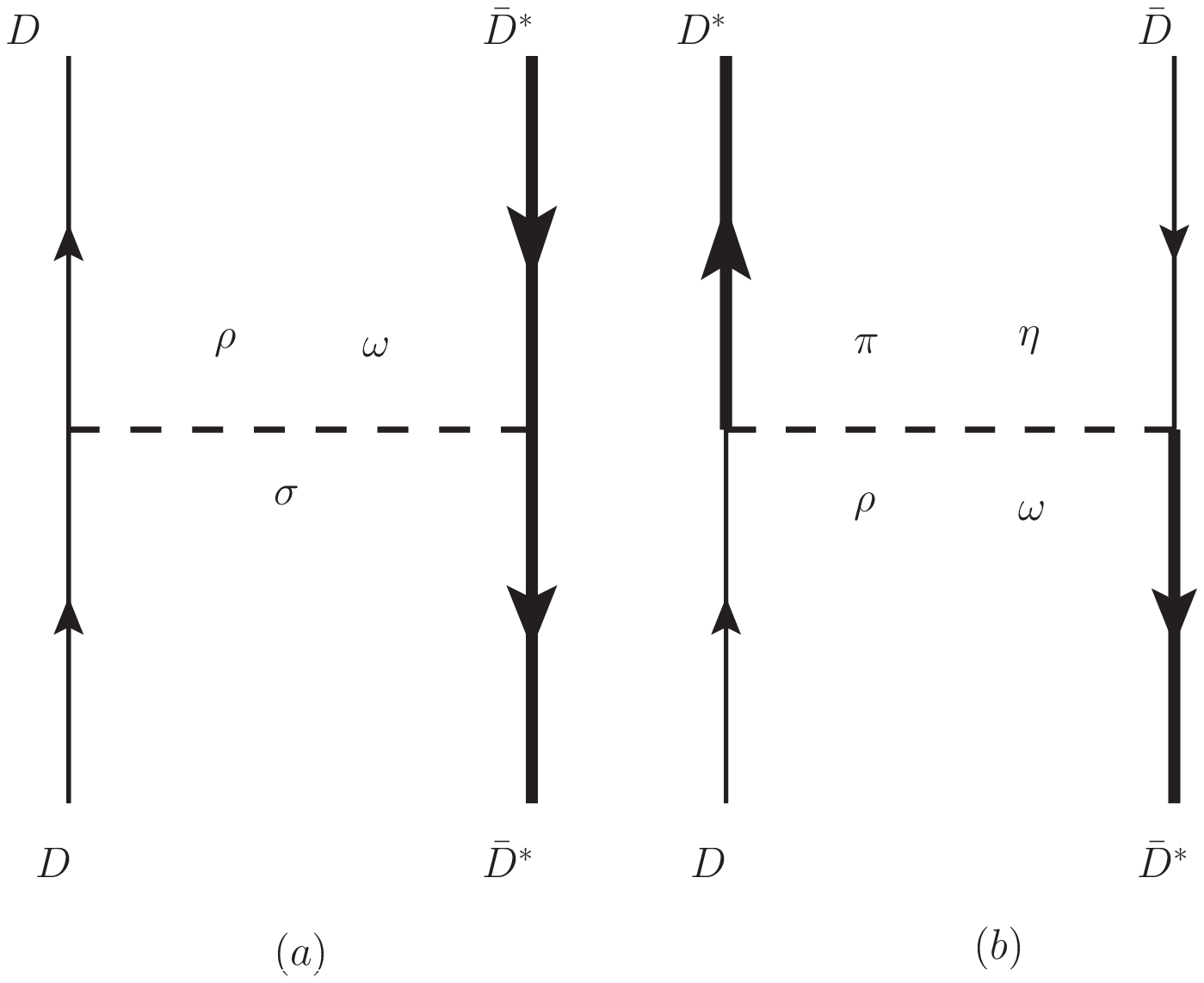}&
\includegraphics[width=0.45\textwidth]{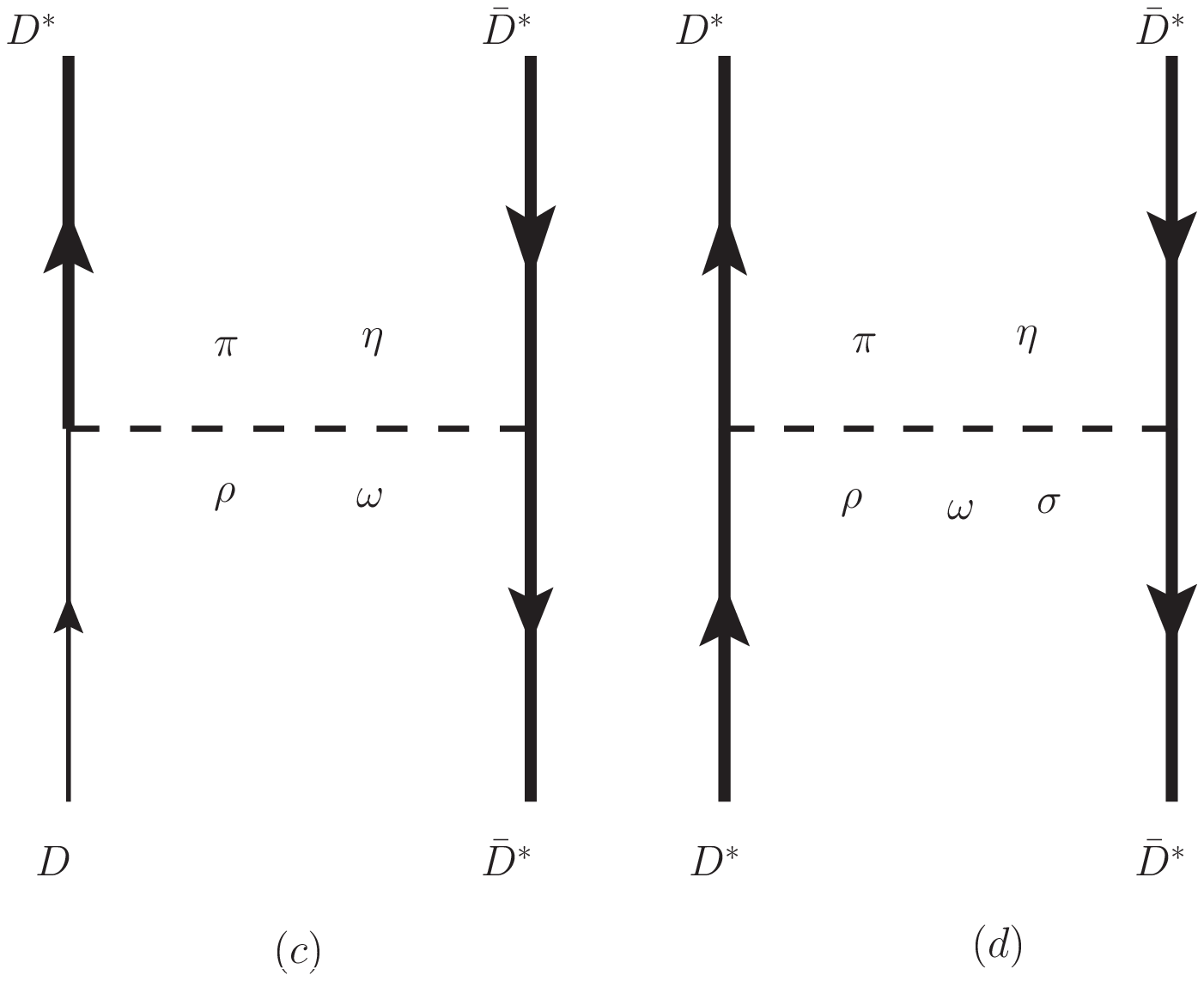}
\end{tabular}
\caption{The feynman diagrams at the tree level. The thick line denotes the heavy
vector meson (or antimeson) while the thin line
stands for the heavy pseudoscalar meson (or antimeson).}\label{Feynmandiagram}
\end{figure}

The are four types of feynman diagrams at the tree level
which are shown in Fig.~\ref{Feynmandiagram}. With the feynman diagrams and the lagrangians
given in Eqs.~(\ref{Lagrangian:p}-\ref{lagrangian:s}),
we derive the effective potentials
with the help of the relation between the effective potential
$V(q)$ and the scattering amplitude $\mathcal{M}(q)$
\begin{eqnarray}
V(q)=-{\mathcal{M}(q)\over \prod\limits_i\sqrt{2M_i}},
\end{eqnarray}
where $M_i$ is the mass of the heavy meson. After the Fourier
transformation, we get the effective potentials in the coordinate
space
\begin{eqnarray}
V(r)={1\over (2\pi)^3}\int d^3{\bf q}e^{i{\bf q}\cdot{\bf
r}}V({\bf q})F^2({\bf q})
\end{eqnarray}
where $F({\bf q})$ is the monopole form factor defined as $F({\bf
q})=(\Lambda^2-m_{ex}^2)/(\Lambda^2-q^2)=(\Lambda^2-m_{ex}^2)/(\chi^2+{\bf
q}^2)$ with $\chi^2=\Lambda^2-q_0^2$. The role of the form factor
is to remove or suppress the contribution from the ultraviolet
region of the exchanged momentum since the light mesons ``see" the
heavy mesons as a whole and do not probe their inner structure.

The expressions of the effective potentials are
\begin{eqnarray}
  V_{\rho/\omega}(r)&=&-C_{\rho/\omega}(i,j){\beta^2g_v^2\over 2} {u\over 4\pi}
  H_0(\Lambda,m_{\rho/\omega},r)S({\bf \epsilon}_4^{\dag}, {\bf \epsilon}_2), \nonumber\\
  V_{\sigma}(r)&=&-C_{\sigma}(i,j)g_s^2{u\over 4\pi}
  H_0(\Lambda,m_{\rho/\omega},r)S({\bf \epsilon}_4^{\dag}, {\bf \epsilon}_2), \label{DDv:DDv}
  \end{eqnarray}
for $D\bar{D}^*\leftrightarrow D\bar{D}^*$,
  \begin{eqnarray}
  V_{\pi}(r)&=&-C_{\pi}(i,j){g^2\over f_{\pi}^2}{\theta^3\over 12\pi}
  \left[M_3(\Lambda,m_{\pi},r)T({\bf \epsilon}_3^{\dag}, {\bf \epsilon}_2)
  +M_1(\Lambda,m_{\pi},r)S({\bf \epsilon}_3^{\dag}, {\bf \epsilon}_2)\right], \nonumber \\
  V_{\eta}(r)&=&-C_{\eta}(i,j){g^2\over f_{\pi}^2}{u^3\over 12\pi}
  \left[H_3(\Lambda,m_{\eta},r)T({\bf \epsilon}_3^{\dag}, {\bf \epsilon}_2)
  +H_1(\Lambda,m_{\eta},r)S({\bf \epsilon}_3^{\dag}, {\bf \epsilon}_2)\right], \nonumber \\
  V_{\rho/\omega}(r)&=&-C_{\rho/\omega}(i,j)\lambda^2g_v^2{u^3\over 6\pi}
  \left[H_3(\Lambda,m_{\rho/\omega},r)T({\bf \epsilon}_3^{\dag}, {\bf \epsilon}_2)
  -2H_1(\Lambda,m_{\rho/\omega},r)S({\bf \epsilon}_3^{\dag}, {\bf \epsilon}_2)\right], \label{DDv:DvD}
  \end{eqnarray}
for $D\bar{D}^*\leftrightarrow D^*\bar{D}$,
  \begin{eqnarray}
  V_{\pi/\eta}(r)&=&C_{\pi/\eta}(i,j){g^2\over f_{\pi}^2}{u^3\over 12\pi}
  \left[H_3(\Lambda, m_{\pi/\eta}, r)T(i{\bf \epsilon}_3^{\dag}\times{\bf \epsilon}_1,
  i{\bf \epsilon}_4^{\dag}\times {\bf \epsilon}_2)
  +H_1(\Lambda, m_{\pi/\eta}, r)S(i{\bf \epsilon}_3^{\dag}\times{\bf \epsilon}_1,
  i{\bf \epsilon}_4^{\dag}\times {\bf \epsilon}_2)\right], \nonumber \\
  V_{\rho/\omega}(r)&=&-C_{\rho/\omega}(i,j){\beta^2g_v^2\over 2}{u\over 4\pi}
  H_0(\Lambda, m_{\rho/\omega}, r)C({\bf \epsilon}_3^{\dag}\cdot {\bf \epsilon}_1,
  {\bf \epsilon}_4^{\dag}\cdot{\bf \epsilon}_2^{\dag}) \nonumber \\
  &&+C_{\rho/\omega}(i,j)\lambda^2g_v^2{u^3\over 6\pi}\left[H_3(\Lambda, m_{\rho/\omega},r)
  T(i{\bf \epsilon}_3^{\dag}\times {\bf \epsilon}_1, i{\bf \epsilon}_4^{\dag}
  \times{\bf \epsilon}_2)
  -2H_1(\Lambda,m_{\rho/\omega},r)S(i{\bf \epsilon}_3^{\dag}\times {\bf \epsilon}_1, i{\bf \epsilon}_4^{\dag}
  \times{\bf \epsilon_2})\right], \nonumber\\
  V_{\sigma}(r)&=&-C_{\sigma}(i,j)g_s^2{u^2\over 4\pi}H_0(\Lambda,m_{\sigma},r)
  C({\bf \epsilon}_3^{\dag}\cdot {\bf \epsilon}_1,
  {\bf \epsilon}_4^{\dag}\cdot{\bf \epsilon}_2^{\dag}), \label{DvDv:DvDv}
  \end{eqnarray}
for $D^*\bar{D}^*\leftrightarrow D^*\bar{D}^*$, and
  \begin{eqnarray}
  V_{\pi/\eta}(r)&=&C_{\pi/\eta}(i,j){g^2\over f_{\pi}^2}{u^3\over 12\pi}
  \left[H_3(\Lambda,m_{\pi/\eta},r)T({\bf \epsilon}_3^{\dag},
  i{\bf \epsilon}_4^{\dag}\times{\bf \epsilon}_2)
  +H_1(\Lambda,m_{\pi/\eta},r)S({\bf \epsilon}_3^{\dag}, i{\bf \epsilon}_4^{\dag}
  \times{\bf \epsilon}_2) \right],\nonumber \\
  V_{\rho/\omega}(r)&=&-C_{\rho/\omega}(i,j)\lambda^2g_v^2{u^3\over 6\pi}
  \left[H_3(\Lambda,m_{\rho/\omega},r)T(i{\bf \epsilon}_3^{\dag}\times
  {\bf \epsilon}_4^{\dag},
  {\bf \epsilon}_2)+H_1(\Lambda,m_{\rho/\omega},r)S(i{\bf \epsilon}_3^{\dag}
  \times {\bf \epsilon}_4^{\dag},
  {\bf \epsilon}_2)\right]\nonumber \\
  &&+C_{\rho/\omega}(i,j)\lambda^2g_v^2{u^3\over 6\pi}
  \left[H_3(\Lambda,m_{\rho/\omega},r)T(i{\bf \epsilon}_3^{\dag}\times {\bf \epsilon}_2,
  {\bf \epsilon}_4^{\dag})+H_1(\Lambda,m_{\rho/\omega},r)
  S(i{\bf \epsilon}_3^{\dag}\times {\bf \epsilon}_2,
  {\bf \epsilon}_4^{\dag})\right], \label{DDv:DvDv}
\end{eqnarray}
for $D\bar{D}^*\leftrightarrow D^*\bar{D}^*$. In the above
equations, $C(A, B)=A B$, $S({\bf A},{\bf B})={\bf A}\cdot {\bf
B}$ and $T({\bf A},{\bf B})=3{\bf A}\cdot\hat{r}{\bf
B}\cdot\hat{r}-{\bf A}\cdot{\bf B}$, which are the generalized
central, spin-spin and tensor operators, respectively.
Their matrix elements are given in
Table~\ref{Matrixelements}.
$C_{\pi/\eta/\sigma/\rho/\omega}(i,j)$ is the channel-dependent
coefficient, and its numerical value is given in Table~\ref{Coefficient}.
 The functions $H_0(\Lambda,m_{ex},r)$,
$H_1(\Lambda,m_{ex},r)$, $H_3(\Lambda,m_{ex},r)$,
$M_1(\Lambda,m_{ex},r)$ and $M_3(\Lambda,m_{ex},r)$ are given in
the APPENDIX. $u^2=m_{ex}^2-q_0^2$ and
$\theta^2=-(m_{\pi}^2-q_0^2)$ with $q_0$ adopted as,

\begin{eqnarray}
D^0\bar{D}^{*0}&\leftrightarrow&D^0\bar{D}^{*0},\quad q_0=0,\qquad
D^0\bar{D}^{*0}\leftrightarrow D^{*0}\bar{D}^0,\quad q_0=m_{D^{*0}}-m_{D^0}, \nonumber \\
D^0\bar{D}^{*0}&\leftrightarrow&D^{+}D^{*-}, \quad q_0=0,\qquad
D^0\bar{D}^{*0}\leftrightarrow D^{*+}D^-,\quad q_0=m_{D^{*0}}-m_{D^0}, \nonumber\\
D^+D^{*-}&\leftrightarrow&D^+D^{*-},\quad q_0=0,\qquad
D^+D^{*-}\leftrightarrow D^{*+}D^-,\quad q_0=m_{D^{*\pm}}-m_{D^{\pm}}, \nonumber\\
D^0\bar{D}^{*0}&\leftrightarrow&D^{*0}\bar{D^{*0}},\quad q_0={m_{D^{*0}}-m_{D^0}\over 2},\qquad
D^0\bar{D}^{*0}\leftrightarrow D^{*+}D^{*-}\quad q_0={m_{D^{*0}}-m_{D^{0}}\over 2}, \nonumber\\
D^+D^{*-}&\leftrightarrow& D^{*0}\bar{D}^{*0},\quad q_0={m_{D^{*\pm}}-m_{D^{\pm}}\over 2},\qquad
D^+D^{*-}\leftrightarrow D^{*+}D^{*-},\quad q_0={m_{D^{*\pm}}-m_{D^{\pm}}\over 2}, \nonumber\\
D^*\bar{D}^*&\leftrightarrow& D^*\bar{D}^*,\quad q_0=0.
\end{eqnarray}
For the pion exchange in the transition process
$D\bar{D}^*\leftrightarrow D^*\bar{D}$, $m_{D^*}-m_D>m_{\pi}$,
which leads to the complex effective potential. Here, we take its
real part ~\cite{Liu:2008fh}, which has a oscillation form as
shown in Eq.~\ref{DDv:DvD}.

\begin{table}[htp]
\renewcommand{\arraystretch}{1.3}
\caption{The matrix elements of the operators appearing in
Eqs.~(\ref{DDv:DDv}-\ref{DDv:DvDv}).}\label{Matrixelements}
\begin{tabular*}{18cm}{@{\extracolsep{\fill}}ccccccc}
\toprule[0.8pt]
\toprule[0.8pt]\addlinespace[3pt]
$\Delta$               &$S({\bf \epsilon^{\dag}_4},{\bf \epsilon_2})$&
                        $T({\bf \epsilon^{\dag}_3},{\bf \epsilon_2})$&
                       $S({\bf \epsilon^{\dag}_3},{\bf \epsilon_2})$ &
$C({\bf \epsilon^{\dag}_3}\cdot{\bf \epsilon_1},{\bf \epsilon^{\dag}_4}\cdot{\bf \epsilon_2})$&
$T(i{\bf \epsilon^{\dag}_3}\times{\bf \epsilon^{\dag}_1},i{\bf \epsilon^{\dag}_4}\times{\bf \epsilon_2})$&
$S(i{\bf \epsilon^{\dag}_3}\times{\bf \epsilon^{\dag}_1},i{\bf \epsilon^{\dag}_4}\times{\bf \epsilon_2})$\\
\specialrule{0.8pt}{3pt}{3pt}
$< ^3S_1|\Delta|^3S_1>$&  1  & 0         &  1   &  1   &  0       &  $-1$  \\
$<^3S_1|\Delta|^3D_1>$&  0  &$-\sqrt{2}$&  0   &  0   &$\sqrt{2}$&  0     \\
$<^3D_1|\Delta|^3S_1>$&  0  &$-\sqrt{2}$&  0   &  0   &$\sqrt{2}$&  0     \\
$<^3D_1|\Delta|^3D_1>$&  1  &    1      &  1   &  1   &   $-1$   &  $-1$  \\
\specialrule{0.8pt}{3pt}{3pt}
$\Delta$       &$T({\bf \epsilon^{\dag}_3},i{\bf \epsilon^{\dag}_4}\times{\bf \epsilon_2})$&
        $S({\bf \epsilon^{\dag}_3},i{\bf \epsilon^{\dag}_4}\times{\bf \epsilon_2})$&
        $T(i{\bf \epsilon^{\dag}_3}\times{\bf \epsilon^{\dag}_4},{\bf \epsilon_2})$&
        $S(i{\bf \epsilon^{\dag}_3}\times{\bf \epsilon^{\dag}_4},{\bf \epsilon_2})$&
        $T(i{\bf \epsilon^{\dag}_3}\times{\bf \epsilon_2},{\bf \epsilon^{\dag}_4})$&
        $S(i{\bf \epsilon^{\dag}_3}\times{\bf \epsilon_2},{\bf \epsilon^{\dag}_4})$\\ \specialrule{0.8pt}{3pt}{3pt}
$<^3S_1|\Delta|^3S_1>$& 0                  &$\sqrt{2}$&  0       &$\sqrt{2}$  &  0                &$-\sqrt{2}$ \\
$<^3S_1|\Delta|^3D_1>$& 1                  & 0        &$-2$      & 0          &$-1$               &  0         \\
$<^3D_1|\Delta|^3S_1>$& 1                  & 0        &$-2$      & 0          &$-1$               &  0         \\
$<^3D_1|\Delta|^3D_1>$&$-{1\over \sqrt{2}}$&$\sqrt{2}$&$\sqrt{2}$&$\sqrt{2}$  &${1\over \sqrt{2}}$&$-\sqrt{2}$  \\
\bottomrule[0.8pt]\bottomrule[0.8pt]
\end{tabular*}
\end{table}

\begin{table}[htp]
\renewcommand{\arraystretch}{1.5}
\caption{The numerical values of the channel-dependent coefficients in
Eqs.~(\ref{DDv:DDv}-\ref{DDv:DvDv}).
$\left[D^{0}\bar{D}^{*0}\right]\equiv {1\over
\sqrt{2}}\left(D^{0}\bar{D}^{*0}-D^{*0}\bar{D}^{0}\right)$,
$\left[D^+D^{*-}\right]\equiv{1\over
\sqrt{2}}\left(D^+D^{*-}-D^{*+}D^-\right)$ and
$\left\{D^*\bar{D}^*\right\}\equiv {1\over
\sqrt{2}}\left(D^{*0}\bar{D}^{*0}+D^{*+}D^{*-}\right)$. For simplicity, we
denote the channel with the form $D\bar{D}^*\leftrightarrow D\bar{D}^*$ as
the ``Direct" channel and the channel with the form $D\bar{D}^*\leftrightarrow D^*\bar{D}$ as
the ``Cross" channel.}\label{Coefficient}
\begin{tabular*}{18cm}{@{\extracolsep{\fill}}ccccccccc}
\toprule[0.8pt]
\toprule[0.8pt]\addlinespace[3pt]
Channels        &        ~        & $C_{\pi^0}$    &  $C_{\pi^{\pm}}$  & $C_{\eta}$       &
 $C_{\rho^0}$   & $C_{\rho^{\pm}}$& $C_{\omega}$   &  $C_{\sigma}$~~ \\
 \specialrule{0.8pt}{3pt}{3pt}
$\left[D^0\bar{D}^{*0}\right]\leftrightarrow\left[D^0\bar{D}^{*0}\right]$&
Direct          &         ~        &       ~        &        ~         &${1\over 2}$     &
         ~      &${1\over 2}$      &        1        \\
      ~         &
Cross           &$-{1\over 2}  $   &       ~        &  $-{1\over 6}$  &$-{1\over 2}$
&        ~      & $-{1\over 2}$    &       ~        \\
$\left[D^0\bar{D}^{*0}\right]\leftrightarrow\left[D^+D^{*-}\right]$ &
Direct          &         ~        &          ~     &        ~        &       ~          &
        1       &         ~        &       ~        \\
       ~        &
      Cross     &       ~         &       $-1$     &         ~        &       ~          &
       $-1$     &       ~         &         ~        \\
$\left[D^+D^{*-}\right]\leftrightarrow\left[D^+D^{*-}\right]$&
Direct          &         ~        &       ~        &        ~         &${1\over 2}$     &
         ~      &${1\over 2}$      &        1        \\
      ~         &
Cross           &$-{1\over 2}  $   &       ~        &  $-{1\over 6}$  &$-{1\over 2}$
&        ~      & $-{1\over 2}$    &       ~        \\
$\left[D^0\bar{D}^{*0}\right]\leftrightarrow\left\{D^{*}\bar{D}^{*}\right\}$ &
        ~       & $-{1\over 2}$    &
 $-1$           &$-{1\over 6}$     &$-{1\over 2}$   & $-1$              &$-{1\over 2}$     &
       ~      \\
$\left[D^+D^{*-}\right]\leftrightarrow\left\{D^*\bar{D}^{*}\right\}$ &
        ~       &$-{1\over 2}$     &
  $-1$          & $-{1\over 6}$    &$-{1\over 2}$   & $-1$              &$-{1\over 2}$       &
       ~      \\
$\left\{D^*\bar{D}^{*}\right\}\leftrightarrow\left\{D^*\bar{D}^{*}\right\}$&
      ~         & ${1\over 2}$     &
       1        &${1\over 6}$      & ${1\over 2}$     &       1          & ${1\over 2}$     &
      1         \\
\bottomrule[0.8pt]\bottomrule[0.8pt]
\end{tabular*}
\end{table}

\section{Numerical Results}\label{numerical}

With the effective potentials given in the
Subsection~\ref{Effect:potential}, we use the FORTRAN program
FESSDE~\cite{Abrashkevich199565,Abrashkevich199890} to solve the
coupled channel Schr\"odinger equation.

\subsection{The Results With The OPE Potential}

Since the hadronic molecule is a loosely bound state composed of
hadrons, one expects that the long-range pion exchange plays a
dominant role among the exchanged mesons. To highlight the
contribution of the pion exchange, we first present the numerical
results in the pion exchange model. Now we have only one free
parameter: the cutoff value. In the deuteron case, the cutoff is
fixed around 1 GeV in order to reproduce the properties of the
deuteron within the same one-boson-exchange model.

We collect the numerical results, which include the binding energy
(B.E.), the root-mean-square radius ($r_{rms}$) and the
probability of the individual channel ($P_i)$ with the pion
exchange potential alone for the Cases I, II, III and IV in
Table~\ref{Numerical:OPE}.

For Case I, we find no binding solutions with the cutoff parameter
around $0.8\sim2.0$ GeV. After adding the charged $D\bar{D}^*$
mode and assuming they are degenerate with the neutral mode, we
obtain a loosely bound state with binding energy 0.32 MeV for the
cutoff parameter being 1.55 GeV. The root-mean-square radius is
4.97 fm. The S wave is dominant, with a probability of $98.81\%$
while the contribution of the D wave is $1.19\%$. When we increase
the cutoff parameter to 1.80 GeV, the binding energy increases to
7.70 MeV, and the root-mean-square radius decreases to 1.36 fm.
Comparison of the results of Case I with those of
Case II indicates that the charged mode of $D\bar{D}^*$ strengthens the attraction
 significantly. This can be easily seen from the following
 simple derivation. If we only consider the neutral $D^0\bar{D}^{*0}$,
 we assume the interaction strength with the pion exchange is
 \begin{eqnarray*}
 \mathcal{V}_{\left[D^0\bar{D}^{*0}\rightarrow D^0\bar{D}^{*0}\right]}=V.
 \end{eqnarray*}
 After adding the charged mode of $D\bar{D}^*$, the interaction strength with the
 exact isospin limit of $I=0$ changes into
 \begin{eqnarray*}
 \mathcal{V}_{\left[D^0\bar{D}^{*0}\rightarrow D^0\bar{D}^{*0}\right]}={V\over 2},\quad
  \mathcal{V}_{\left[D^+D^{*-}\rightarrow D^+D^{*-}\right]}={V\over 2},\quad
  \mathcal{V}_{\left[D^0\bar{D}^{*0}\rightarrow D^+D^{*-}\right]}=V,\quad
  \mathcal{V}_{\left[D^+D^{*-} \rightarrow D^0\bar{D}^{*0}\right]}=V.
 \end{eqnarray*}
The total interaction is $3V$, three times of that with only the
neutral $D^0\bar{D}^{*0}$. Actually, this has been pointed out by Close {\it et al}
previously~\cite{Thomas:2008ja}.
Therefore the charged mode of $D\bar{D}^*$ is important
in the formation of a bound state, although the required cutoff
parameter is larger than 1.5 GeV. This is consistent with the conclusion of
Ref.~\cite{Aceti:2012cb}, but somewhat different from that of
Refs.~\cite{Artoisenet:2010va,Braaten:2007ft}.
In~\cite{Artoisenet:2010va,Braaten:2007ft}, Braaten {\it et al} studied the line
sharp of $X(3872)$. They conclude
that at energies within only a few MeV of the $D^0\bar{D}^{*0}$ threshold, the results with
 only the neutral $D^0\bar{D}^{*0}$ is accurate but generalized to
 the entire $D\bar{D}^{*}$ threshold the charged $D^+D^{*-}$ plays a significant role.

In Case III, we can see the significant role of the
coupled-channel effects after we turn on the coupling of
$D\bar{D}^*$ to $D^*\bar{D}^*$. In fact, the binding energy
increases by several tens of MeV compared with Case II with the same
cutoff parameter as shown in Table~\ref{Numerical:OPE}. The
binding energy and the root-mean-square radius of the bound state
are 0.76 MeV and 3.79 fm respectively with the cutoff parameter
around 1.10 GeV, which is a reasonable value. The dominant channel
is still ${1\over
2}\left[\left(D^0\bar{D}^{*0}-D^{*0}\bar{D}^0\right)
+\left(D^+D^{*-}-D^{*+}D^-\right)\right]|^3S_1>$, with a
probability of $97.82\%$. The probability of $D^*\bar{D}^*$ is
small, about $(1.24+0.20)\%=1.44\%$.

Since the state in Case I only contains the neutral $D\bar{D}^*$
mode, it is an equal superposition state of the isoscalar and
isovector state. The states in Cases II and III are definitely
isoscalar. Actually, none of the states in Cases I, II and III
correspond to the physical state of $X(3872)$. As mentioned
before, the hidden-charm di-pion decay mode of $X(3872)$ violates
isospin symmetry.

In order to reproduce the physical $X(3872)$ state, we move on to
Case IV and explicitly consider the mass splitting of the
neutral and charged $D(D^\ast)$ mesons. Now the binding energy
decreases by roughly $2.5\sim 3$ MeV compared to Case III with the
same cutoff parameter as shown in Table~\ref{Numerical:OPE}, which
is an expected result since the charged $D^+D^{*-}$ pair is almost
8 MeV heavier than the $D^0\bar{D}^{*0}$ pair. For example, the
binding energy is 0.26 MeV when the cutoff parameter is 1.15 MeV.
For comparison, the binding energy is 2.72 MeV in Case III with
$\Lambda = 1.15$ MeV. We will show below that the flavor wave
function of this very loosely bound molecular state contains a
large isovector component, which decays into the $J\psi \rho$
mode. In other words, this molecular state can be interpreted as
X(3872).

\begin{table}[htp]
\renewcommand{\arraystretch}{1.3}
\caption{The binding solutions of $X(3872)$ with the OPE potential.
$\Lambda$ is the cutoff parameter. ``B.E." is the binding energy
while ``Mass" is the calculated mass of $X(3872)$. $r_{rms}$ and
``$P_i$" are the root-mean-square radius and the probability of
the ith channel, respectively. ``$\times$" means no binding
solutions, and ``$-$" denotes that the corresponding component
does not exist.}\label{Numerical:OPE}
  \begin{tabular*}{18cm}{@{\extracolsep{\fill}}cccccccccccl}
  \toprule[0.8pt]
  \toprule[0.8pt]\addlinespace[3pt]
    Cases         &$\Lambda$(GeV)& B.E. (MeV)& Mass (MeV)&$r_{rms}$ (fm)& $P_1(\%)$&
    $P_2(\%)$     & $P_3(\%)$    & $P_4(\%)$ &$P_5(\%)$  & $P_6(\%)$ \\
    \specialrule{0.6pt}{3pt}{3pt}
    \multirow{3}*{I}&            &           &          &           &         &
        ~         &    $-$       & $-$       & $-$      &  $-$     \\
        ~         &$0.80\sim2.0$ &\multicolumn{5}{c}{$\times$}
                  &    $-$       &  $-$      & $-$      &   $-$    \\
        ~         &              &           &          &           &         &
        ~         &    $-$       & $ -$      & $ - $    &   $-$  \\ [5pt]
\multirow{5}*{II} &    1.55      & $0.32$    & 3871.49  &   4.97
& 98.81   &
        1.19      & $ -$         & $ - $     &  $-$     &   $-$  \\
        ~         &    1.60      & $0.92$    & 3870.89  &   3.51   & 98.39   &
        1.61      &    $ -$      & $ - $     &  $-$     &  $-$   \\
        ~         &    1.65      & $1.90$    & 3869.91  &   2.56   & 98.01   &
        1.99      &    $ -$      & $ - $     &  $-$     &  $-$   \\
        ~         &    1.70      & $3.31$    & 3868.50  &   1.99   & 97.69   &
        2.31      &   $ -$       & $ - $     &  $-$     &   $-$  \\
        ~         &    1.80      & $7.70$    & 3864.11  &   1.36   & 97.18   &
        2.82      &    $ -$      & $ - $     & $-$      &  $-$   \\  [5pt]
\multirow{6}*{III}&    1.10      & $0.76$    & 3871.05  &   3.79
& 97.82    &
        0.73      & $-$          & $-$       &    1.24  &   0.20 \\
         ~        &    1.15      & $2.72$    & 3869.09  &   2.17   & 96.15   &
        0.82      &  $-$         & $-$       &  2.64    &   0.40 \\
        ~         &    1.20      & $6.25$    & 3865.56  &   1.49   & 94.26   &
        0.77      &    $-$       &  $-$      &    4.37  &   0.60 \\
        ~         &    1.25      & $11.66$   & 3860.15  &   1.13   & 92.20   &
        0.67      &   $-$        &  $-$      &  6.32    &   0.81 \\
        ~         &    1.30      & $19.21$   & 3852.60  &   0.91   & 90.05   &
        0.55      &   $-$        &  $-$      & 8.38     &   1.02 \\
                  &    1.55      &  95.79    &  3776.02 &   0.47   &  80.68  &
        0.16      &    $-$       &   $-$     &  17.37   &   1.80 \\   [5.0pt]
\multirow{5}*{IV(Phy)} &    1.15      & $0.26$    &  3871.55 &
4.79   &  85.68  &
   0.22           &    12.29     &  0.24     &   0.36   &  0.21  \\
       ~          &    1.17      & $1.03$    &  3870.78 &   2.99   &  76.37  &
  0.30            &    20.27     &  0.33     &   2.39   &  0.35  \\
       ~          &    1.20      & $2.93$    &  3868.88 &   1.84   &  66.18  &
  0.34            &   28.74      &  0.36     &   3.84   &  0.54  \\
       ~          &    1.25      & $7.99$    &  3863.82 &   1.20   &  56.72  &
  0.32            &    35.76     &  0.34     &   6.08   &  0.79  \\
       ~          &     1.30     & $15.36$   &  3856.45 &   0.93   &  51.59  &
  0.27            &    38.61     &  0.28     &   8.25   &  1.01  \\ [3pt]
    \bottomrule[0.8pt]
    \bottomrule[0.8pt]
\end{tabular*}
\end{table}

\subsection{The Results With The OBE Potential}

\begin{figure}[htp]
  \begin{tabular}{ccc}
  \includegraphics[width=0.33\textwidth]{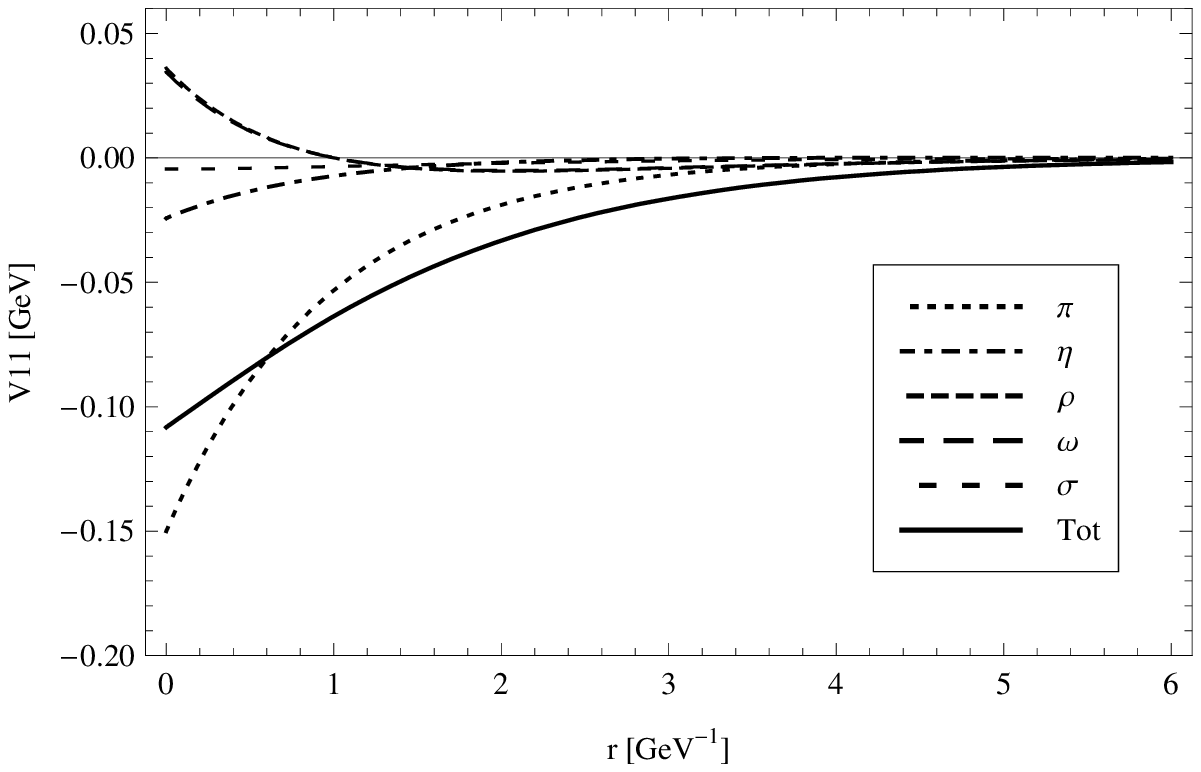}&
  \includegraphics[width=0.33\textwidth]{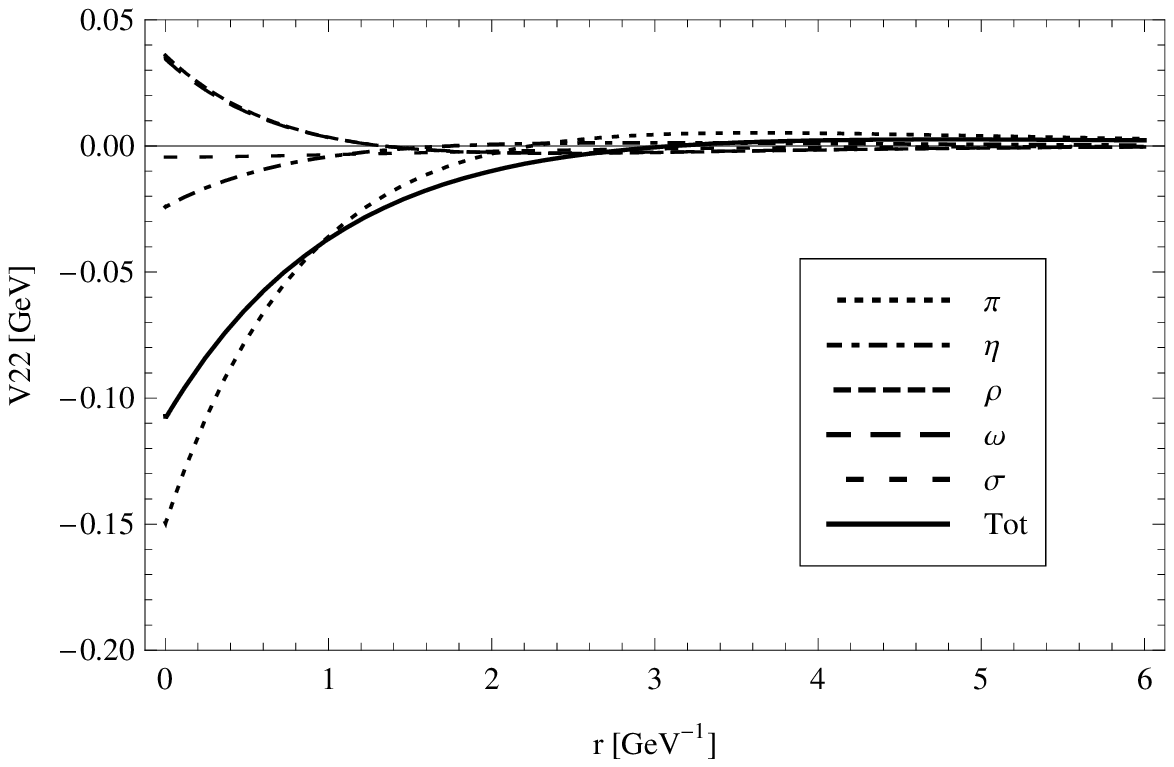}&
  \includegraphics[width=0.33\textwidth]{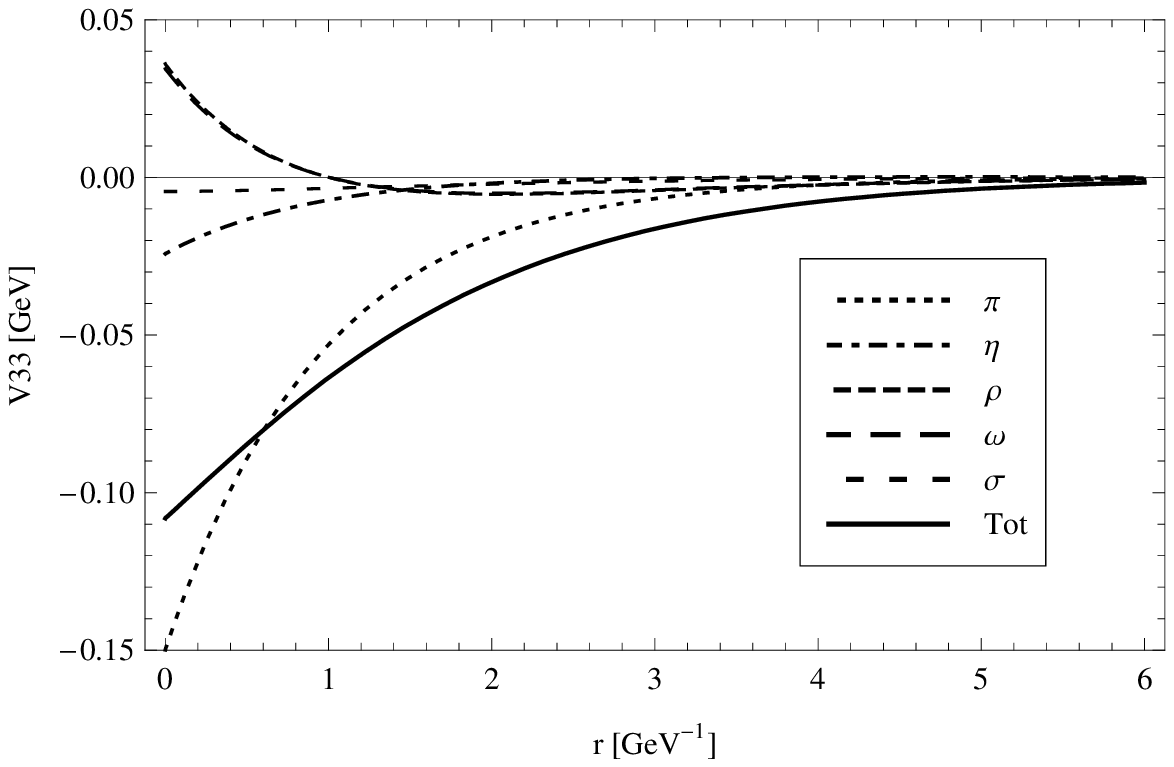}\\
   ($V_{11}$)&($V_{22}$)&($V_{33}$)\\
  \includegraphics[width=0.33\textwidth]{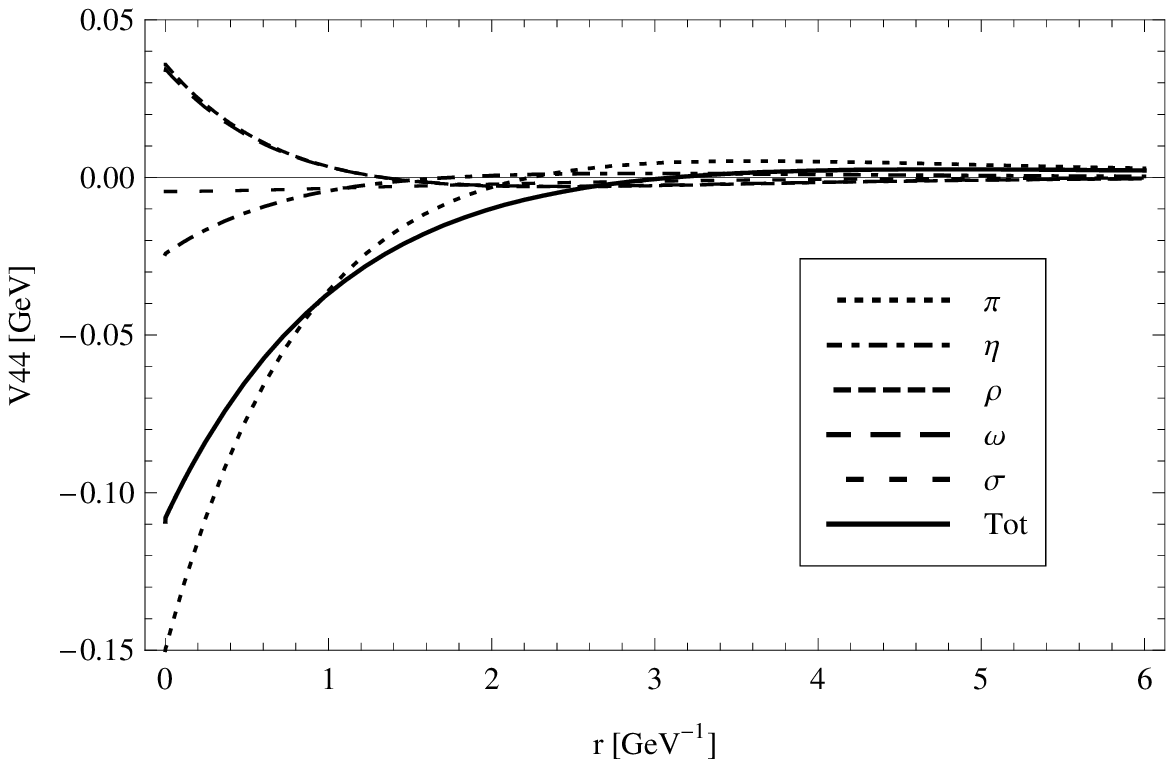}&
  \includegraphics[width=0.33\textwidth]{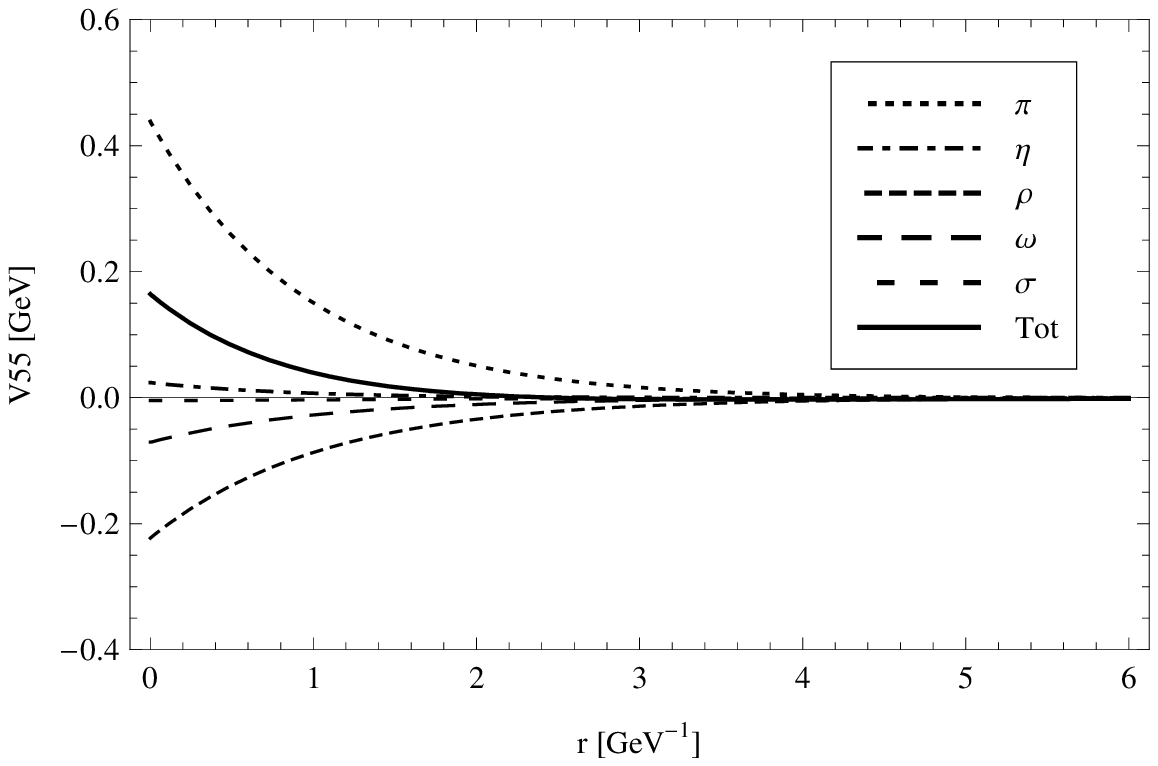}&
  \includegraphics[width=0.33\textwidth]{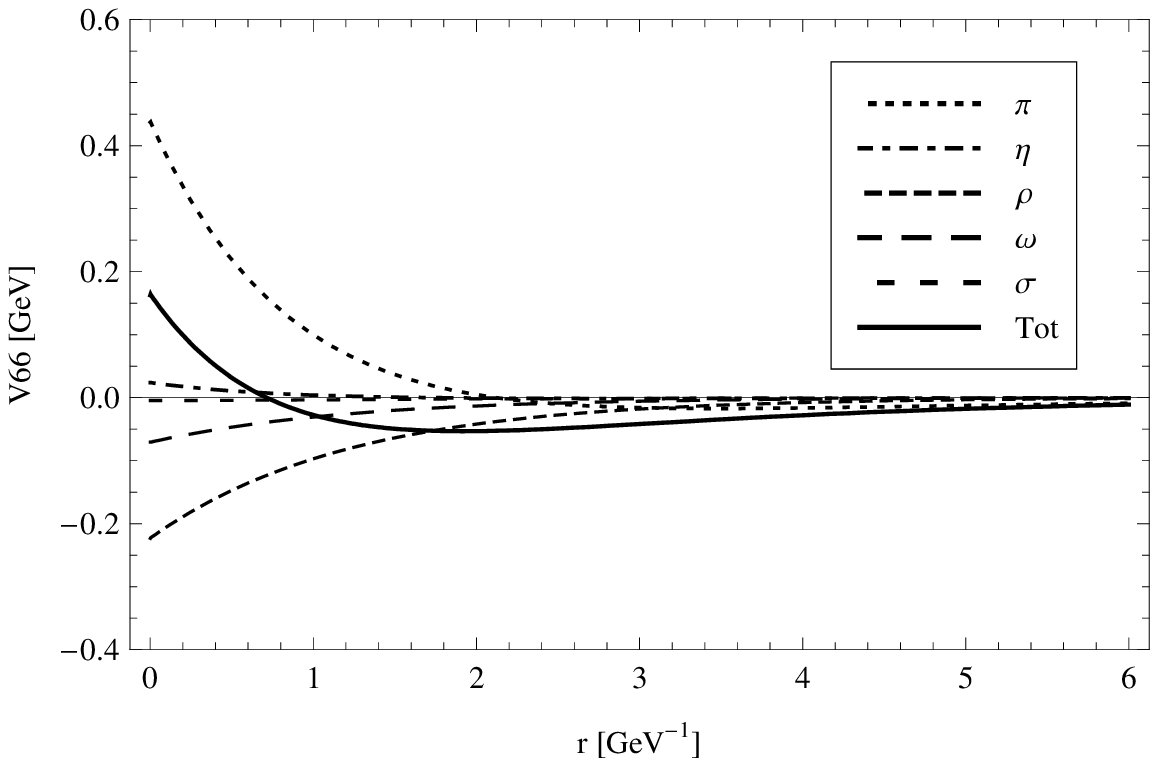}\\
   ($V_{44}$)&($V_{55}$)&($V_{66}$)
  \end{tabular}
  \caption{The potentials for the different channels of $X(3872)$ with $J^{PC}=1^{++}$
  when the cutoff parameter is fixed at 1.05 GeV.}\label{plot:diagonal}
\end{figure}

\begin{figure}[htp]
   \begin{tabular}{ccc}
   \includegraphics[width=0.33\textwidth]{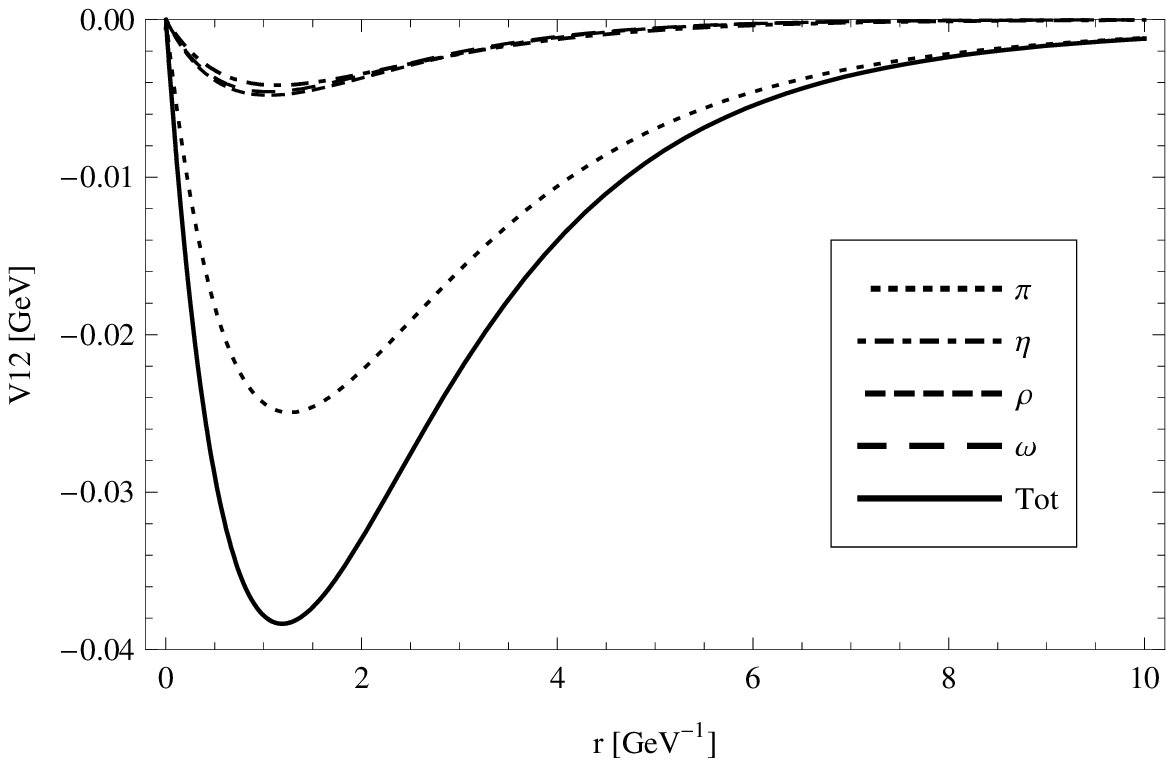}&
   \includegraphics[width=0.33\textwidth]{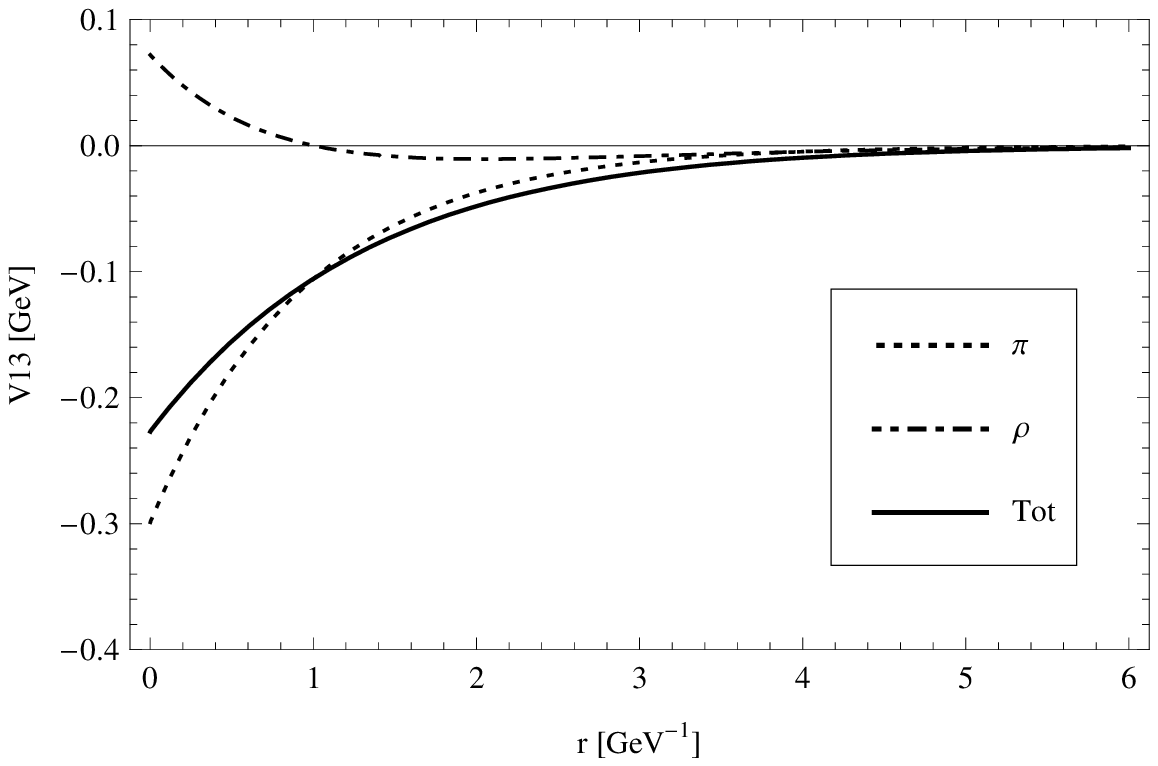}&
   \includegraphics[width=0.33\textwidth]{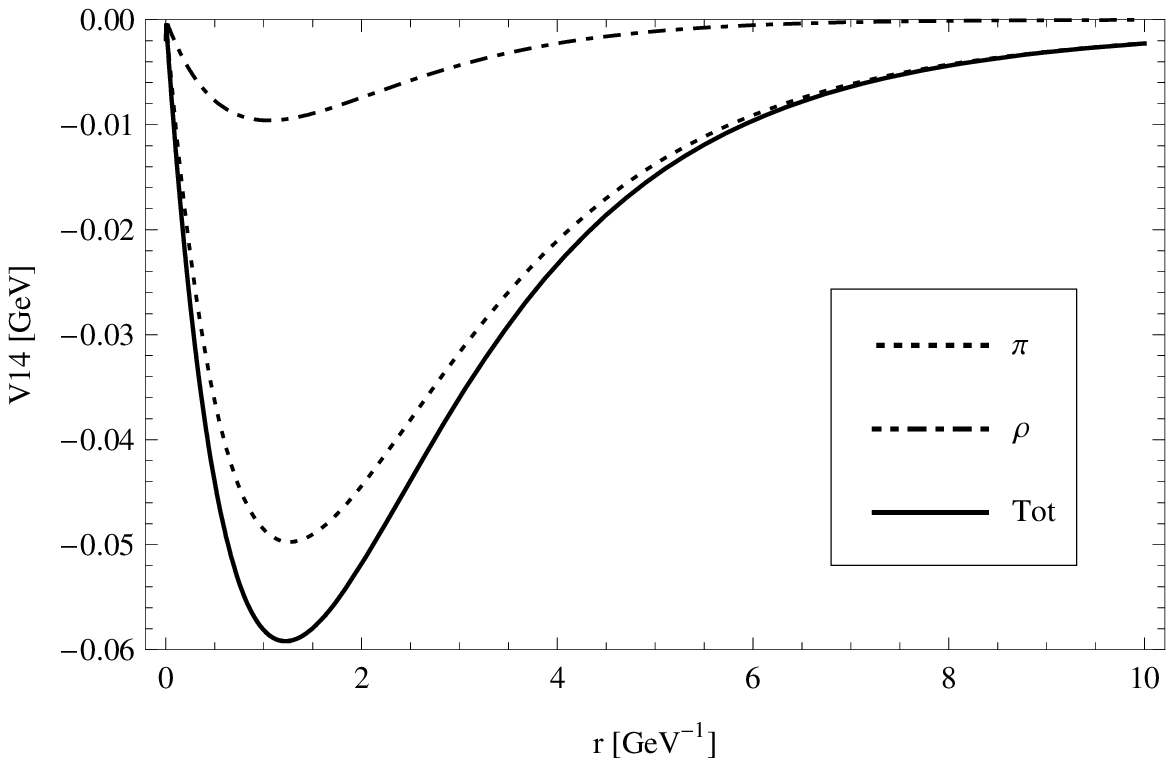}\\
   ($V_{12}$)&($V_{13}$)&($V_{14}$)\\
   \includegraphics[width=0.33\textwidth]{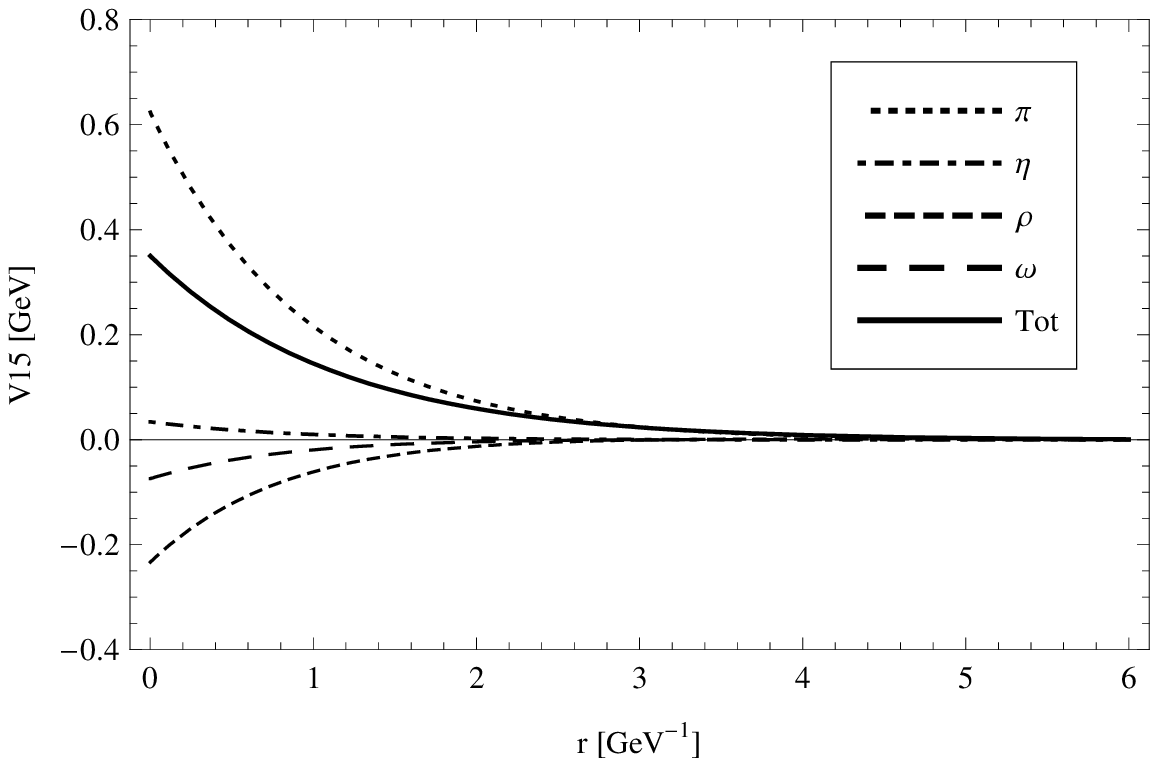}&
   \includegraphics[width=0.33\textwidth]{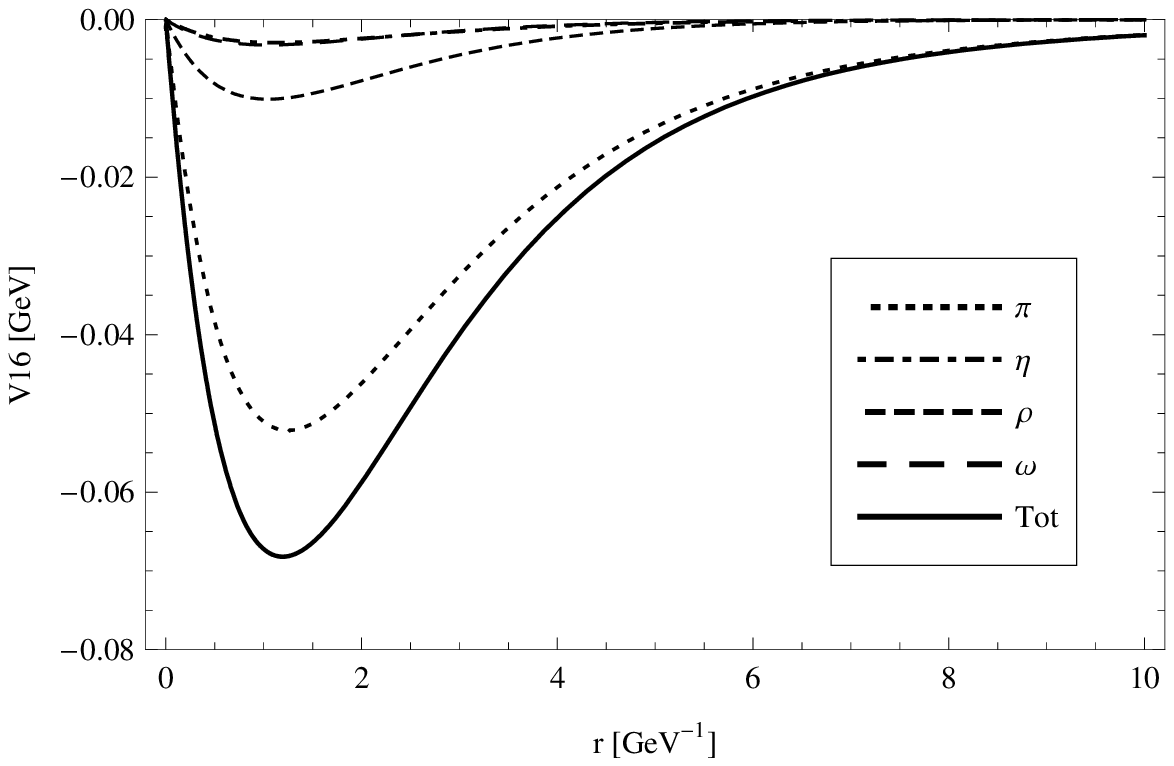}&
   \includegraphics[width=0.33\textwidth]{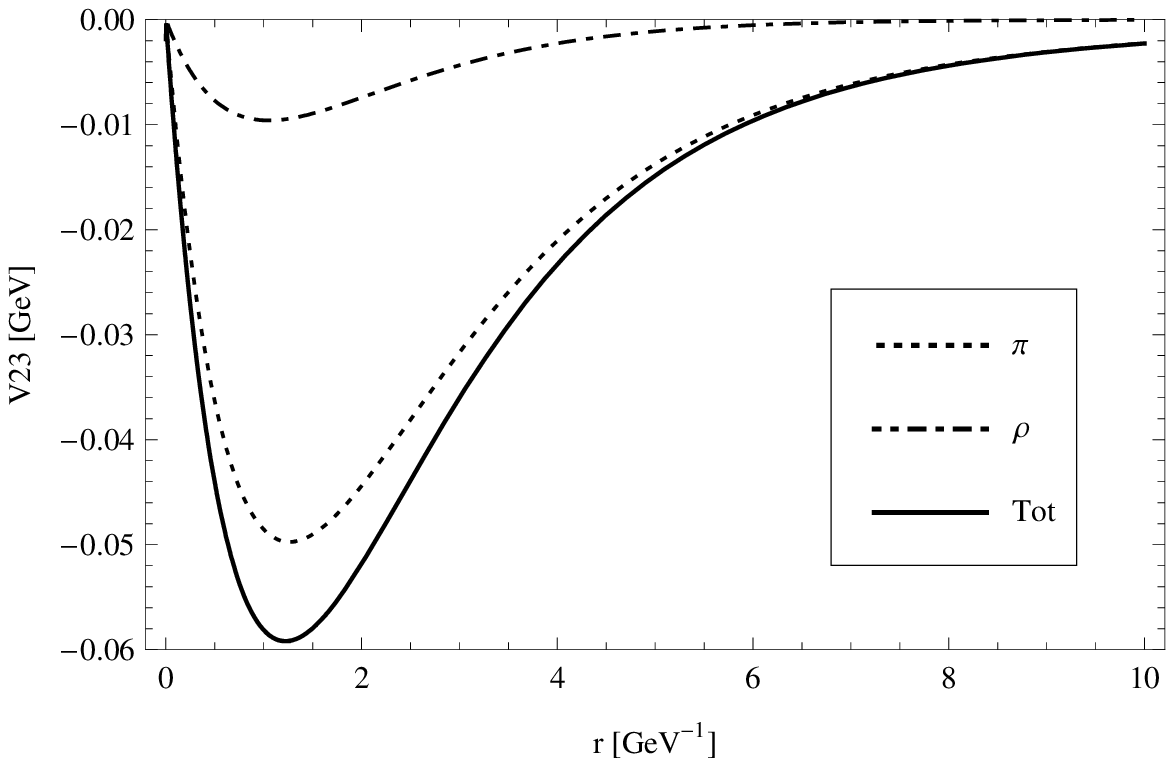}\\
   ($V_{15}$)&($V_{16}$)&($V_{23}$)\\
   \includegraphics[width=0.33\textwidth]{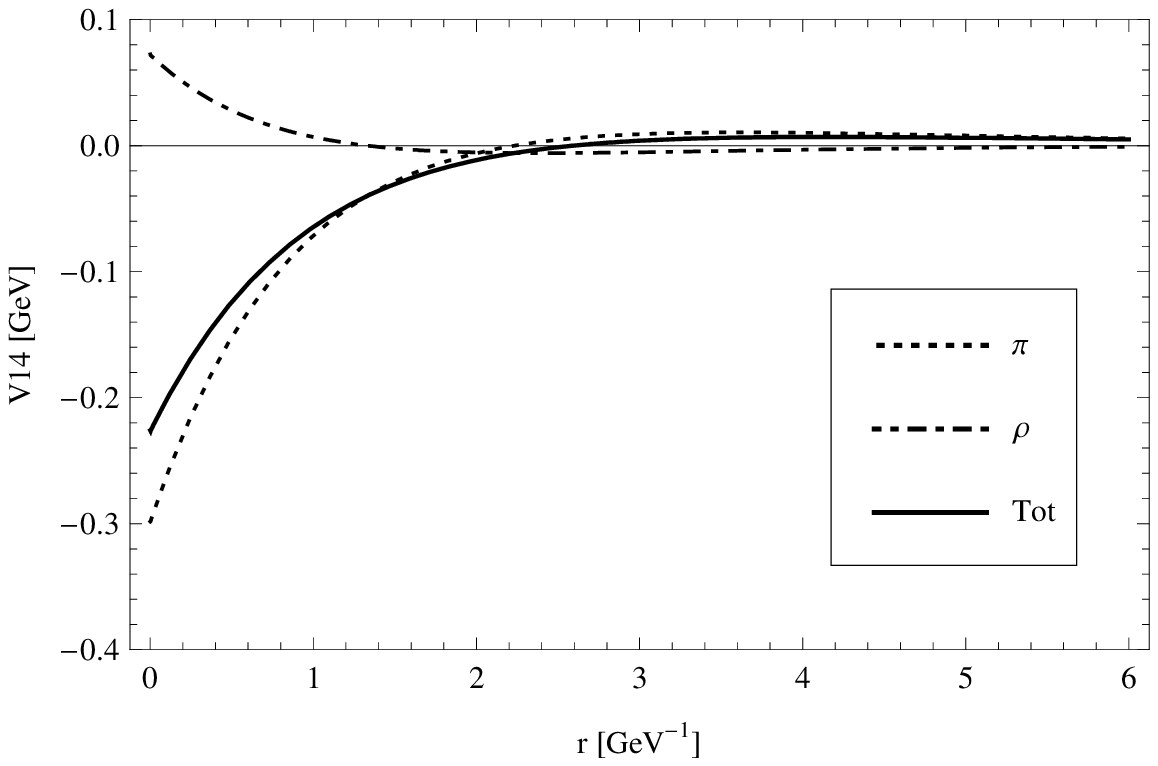}&
   \includegraphics[width=0.33\textwidth]{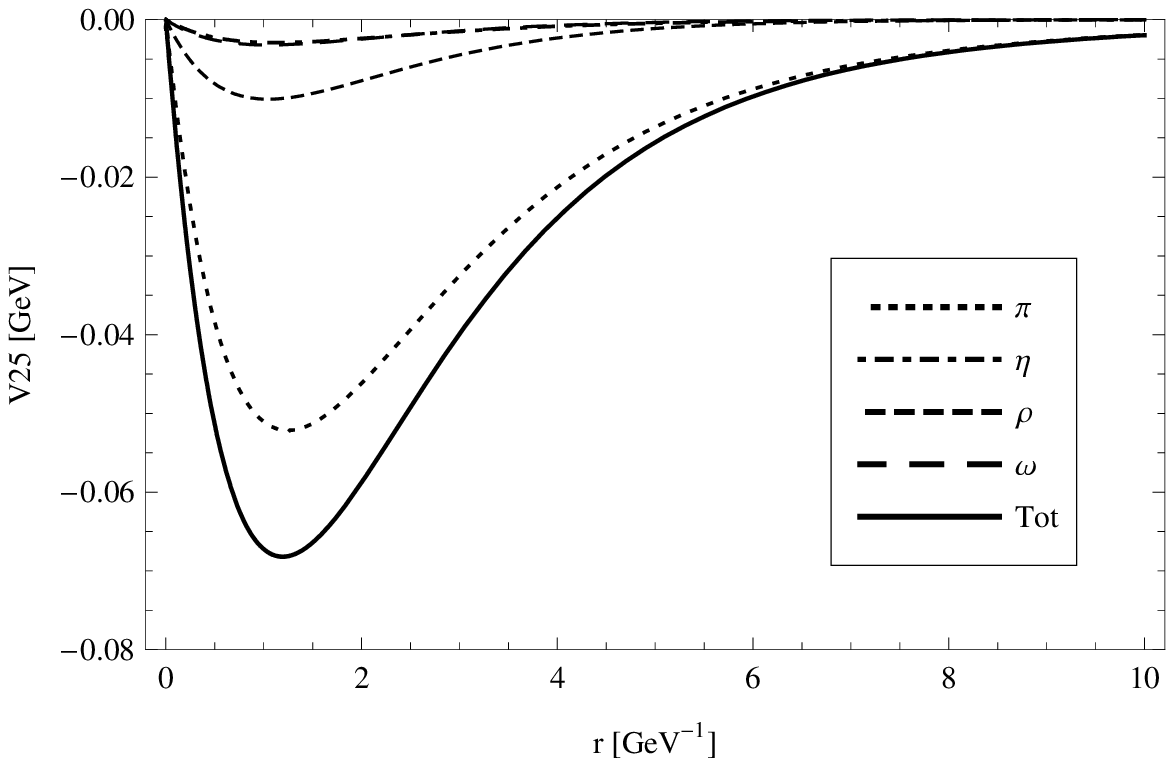}&
   \includegraphics[width=0.33\textwidth]{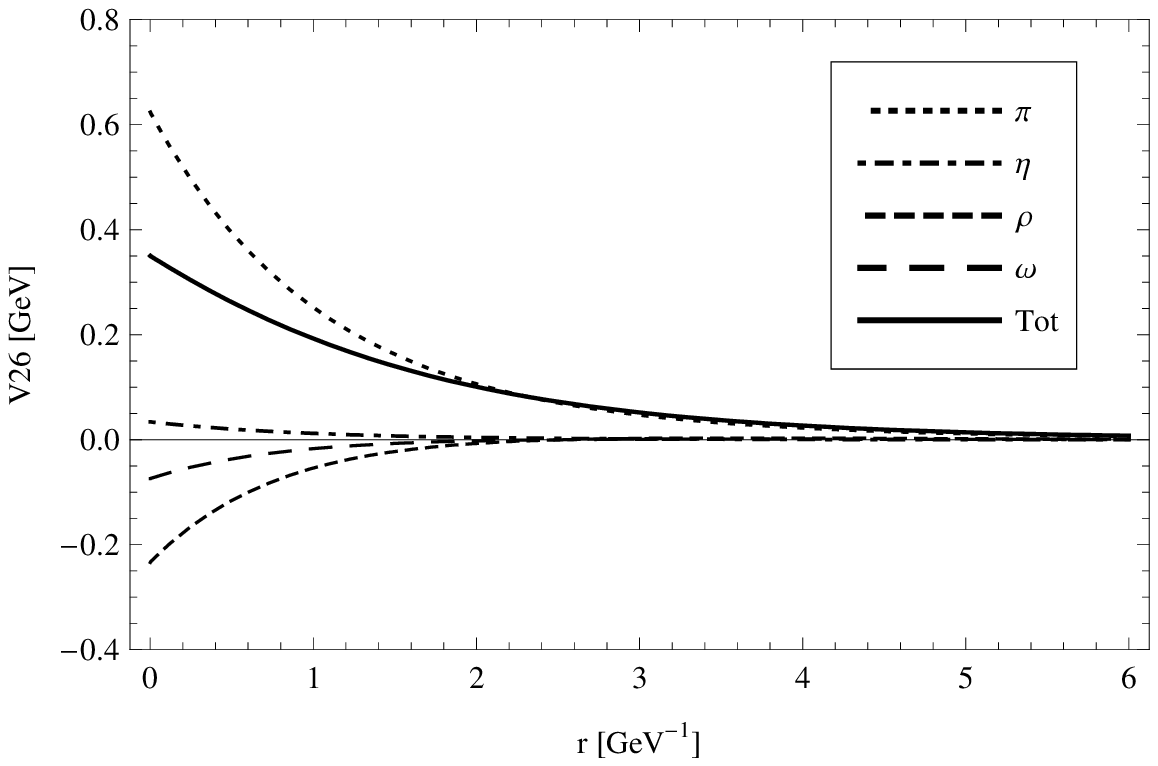}\\
   ($V_{24}$)&($V_{25}$)&($V_{26}$)\\
   \includegraphics[width=0.33\textwidth]{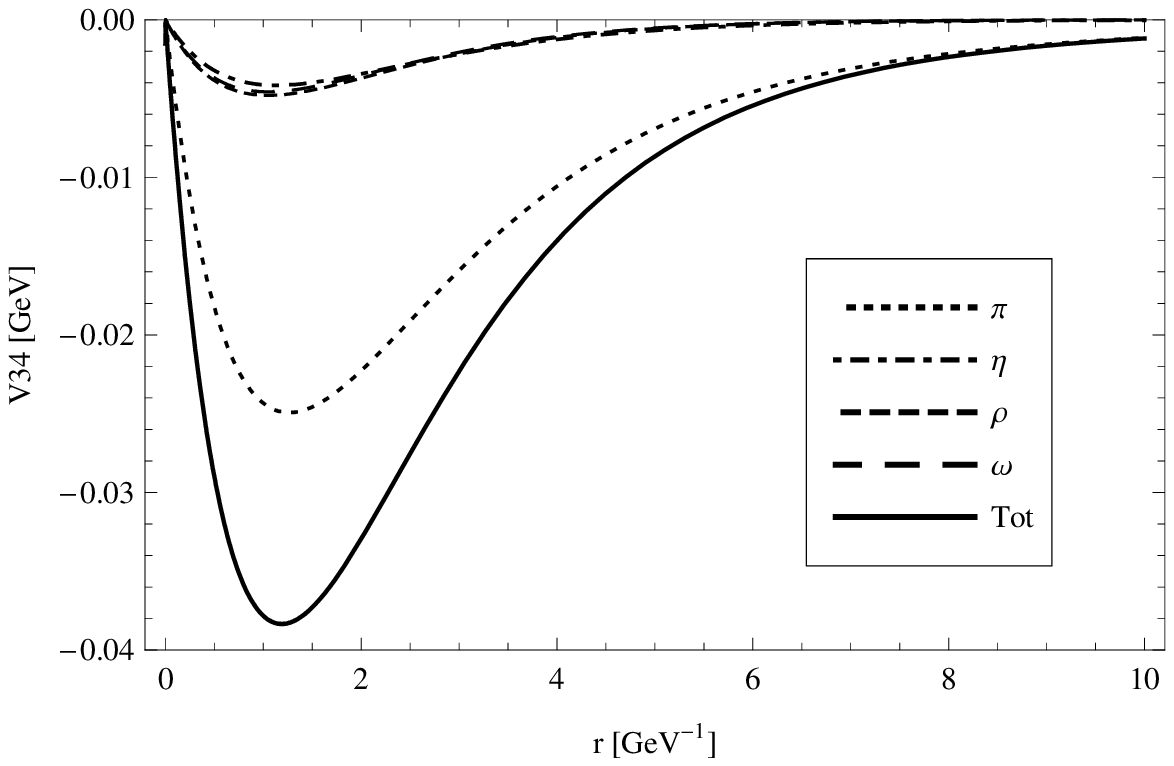}&
   \includegraphics[width=0.33\textwidth]{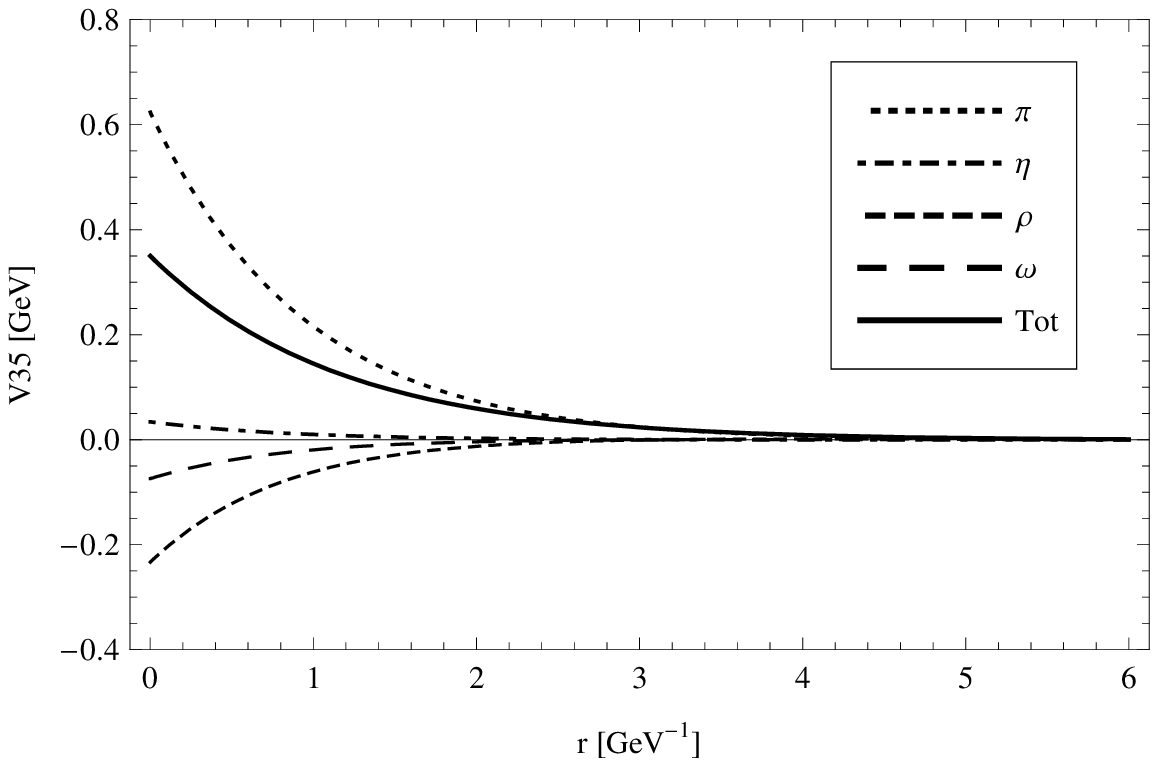}&
   \includegraphics[width=0.33\textwidth]{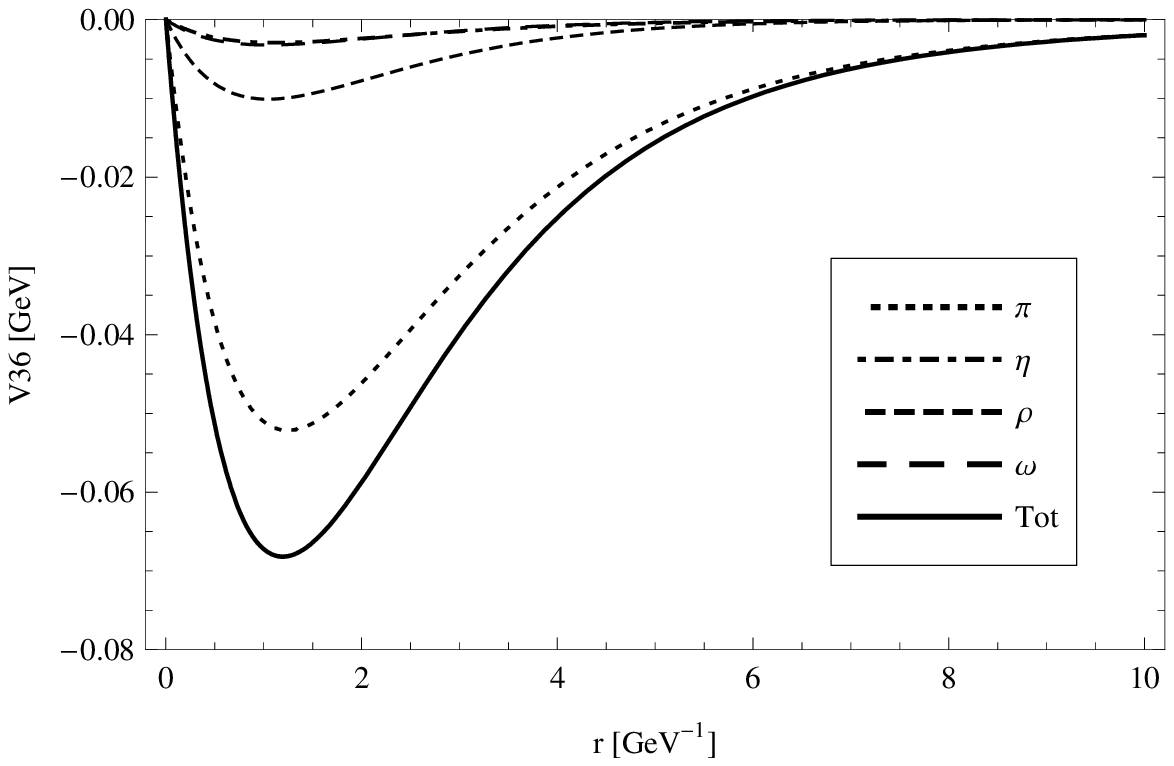}\\
   ($V_{34}$)&($V_{35}$)&($V_{36}$)\\
   \includegraphics[width=0.33\textwidth]{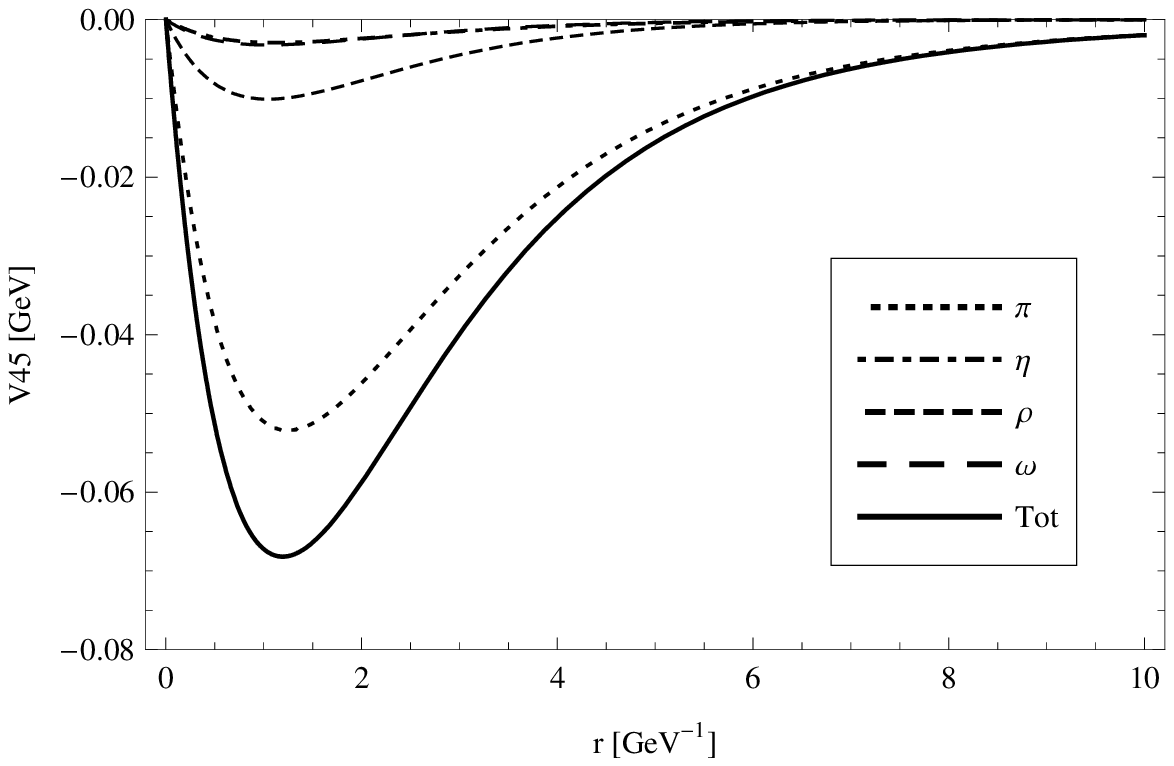}&
   \includegraphics[width=0.33\textwidth]{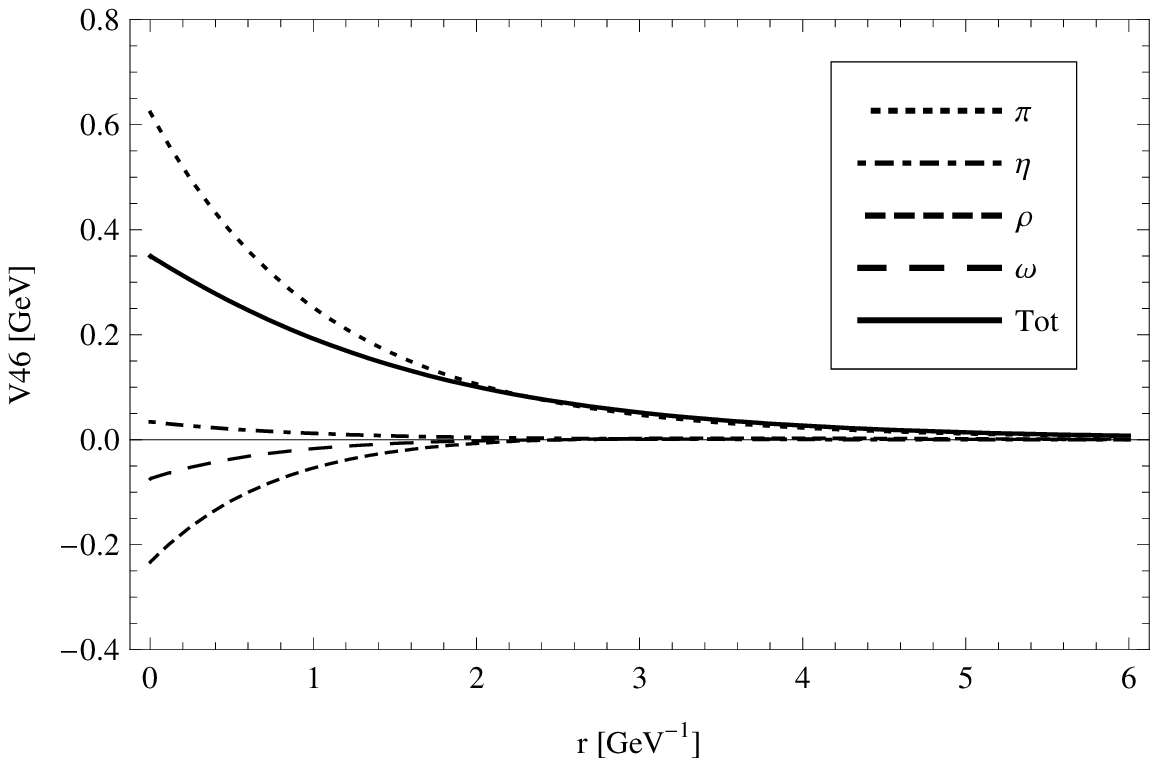}&
   \includegraphics[width=0.33\textwidth]{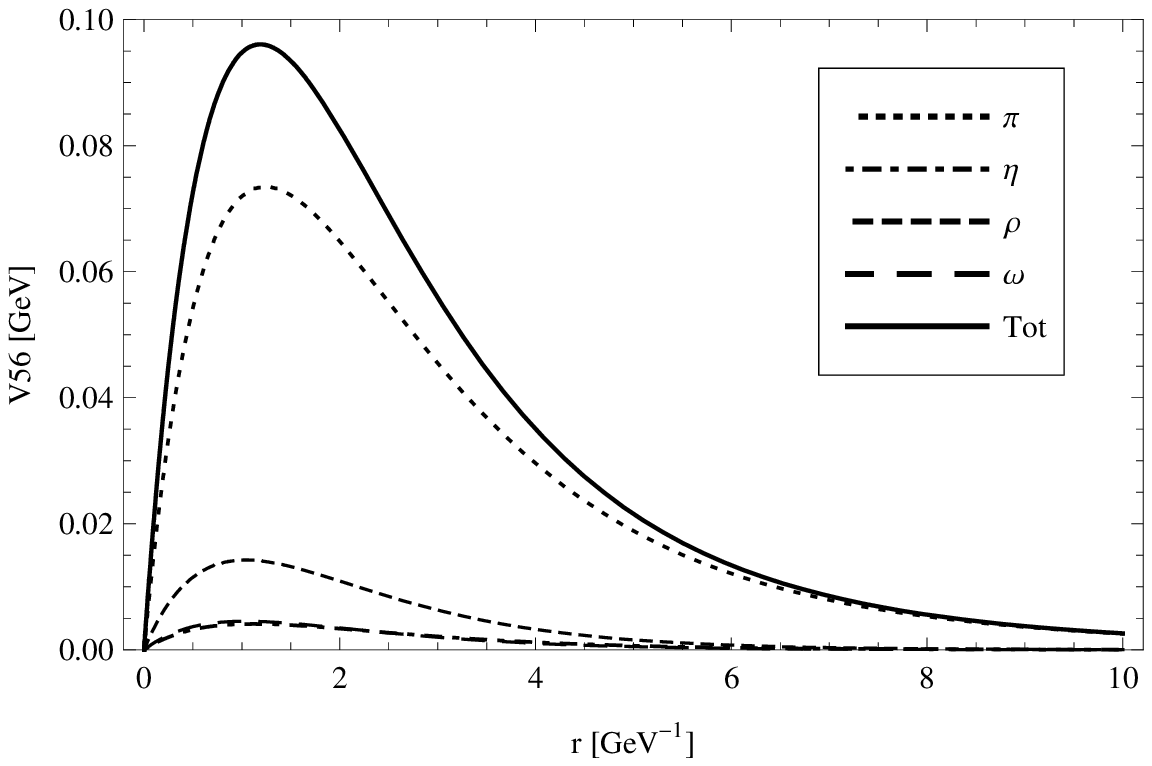}\\
   ($V_{45}$)&($V_{46}$)&($V_{56}$)
   \end{tabular}
   \caption{The potentials of the different channels of X(3872) with $J^{PC}=1^{++}$
   when the cutoff parameter is fixed at 1.05 GeV.}\label{plot:nondiagonal}
\end{figure}

Taking into account the heavier $\eta$, $\sigma$, $\omega$ and
$\rho$ exchanges as well as the pion exchange, we collect the
numerical results for Cases I, II, III and IV with the OBE
potential in Table~\ref{Numerical:OBE}. To make a rough estimation
of the specific role of the exchanged meson, we plot the effective
potential for Case IV when the cutoff parameter is fixed at 1.05
GeV in Figs.~\ref{plot:diagonal} and~\ref{plot:nondiagonal}. From
Figs.~\ref{plot:diagonal} and~\ref{plot:nondiagonal}, we know that
the heavier $\eta$, $\sigma$, $\omega$ and $\rho$ exchange cancel
each other to a large extent. Therefore, the pion exchange plays a
dominant role in forming the loosely bound state. Although the
potentials of the heavier scalar and vector meson exchange cancel
each other greatly, the residual effect of the heavier meson
exchange can still modify the binding solution.

Different from the OPE case, we obtain a loosely bound state with
binding energy 0.21 MeV and root-mean-square radius 5.36 fm with
the cutoff parameter around 1.85 GeV in Case I. In other words,
the heavier scalar and vector meson exchange plays some role in
the formation of the bound state. In Case II, if the cutoff
parameter is fixed at 1.10 GeV, the binding energy and the
root-mean-square radius of the bound state obtained are 0.61 MeV
and 4.21 fm, respectively. With the OBE potential, the coupling of
$D\bar{D}^*$ to $D^*\bar{D}^*$ increases the binding energy by
about 5 MeV. For example, if the $D^*\bar{D}^*$ component is not
included, the binding energy is 0.61 MeV for the cutoff parameter
fixed at 1.10 GeV. In contrast, after turning on the coupling of
$D\bar{D}^*$ to $D^*\bar{D}^*$, the binding energy increases to
5.69 MeV with the same cutoff, see Cases II and III in
Table~\ref{Numerical:OBE}.

\begin{figure}[htp]
\centering
\includegraphics[width=0.6\textwidth]{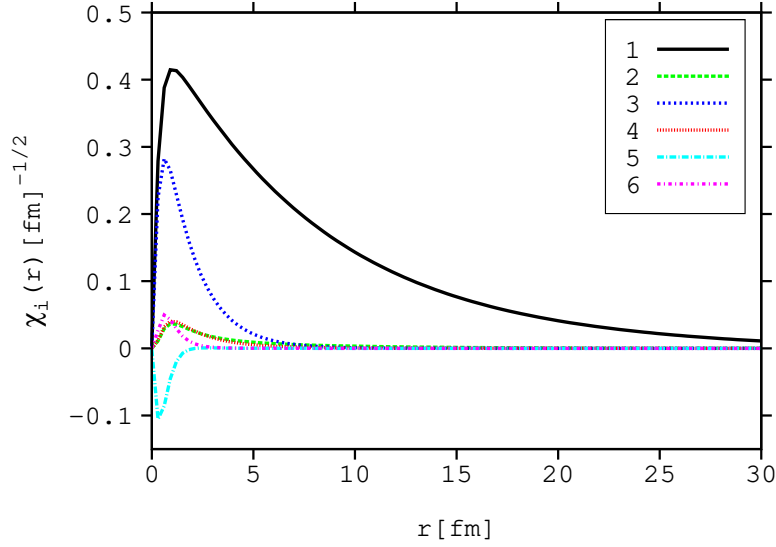}
\caption{(Color online) The radial wave functions of the different
channels of $X(3872)$ when the binding energy is 0.30
MeV.}\label{Plot:Wave}
\end{figure}

\begin{figure}[htp]
\begin{tabular}{cc}
\includegraphics[width=0.45\textwidth]{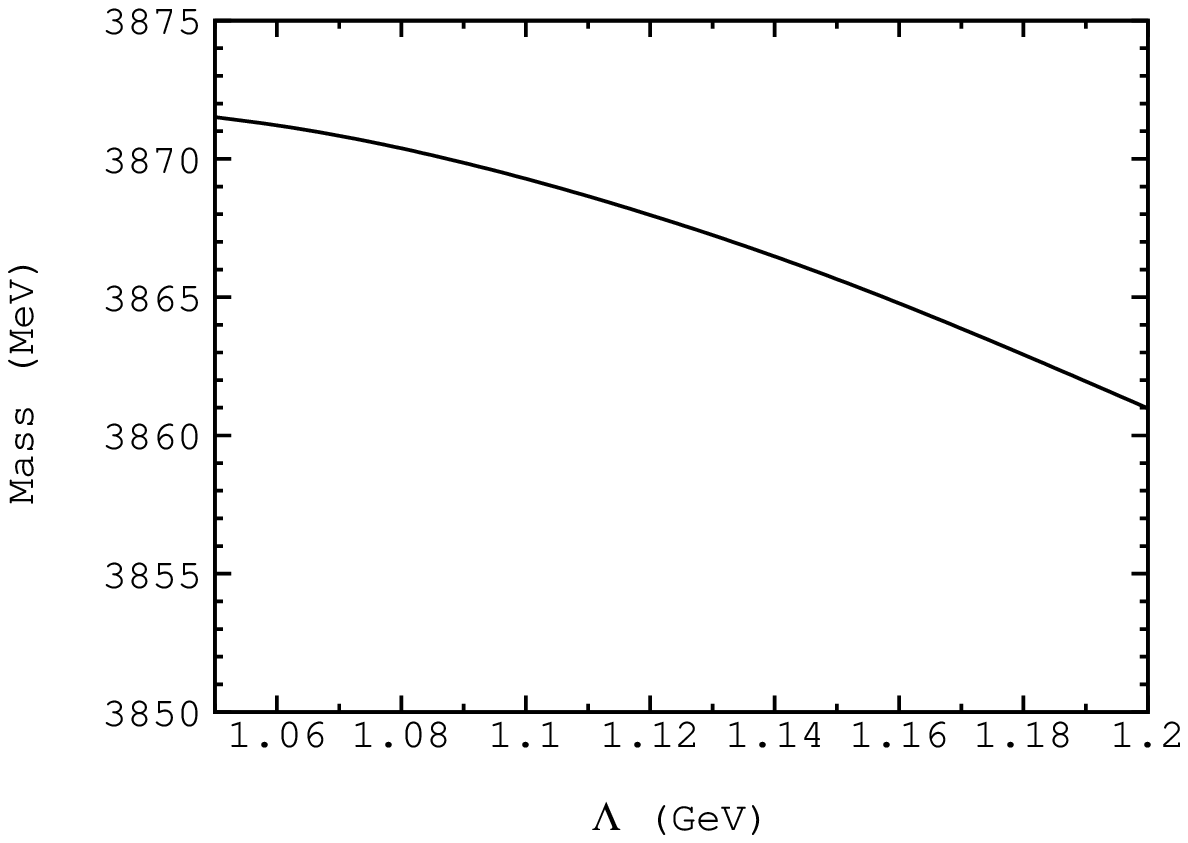}&
\includegraphics[width=0.45\textwidth]{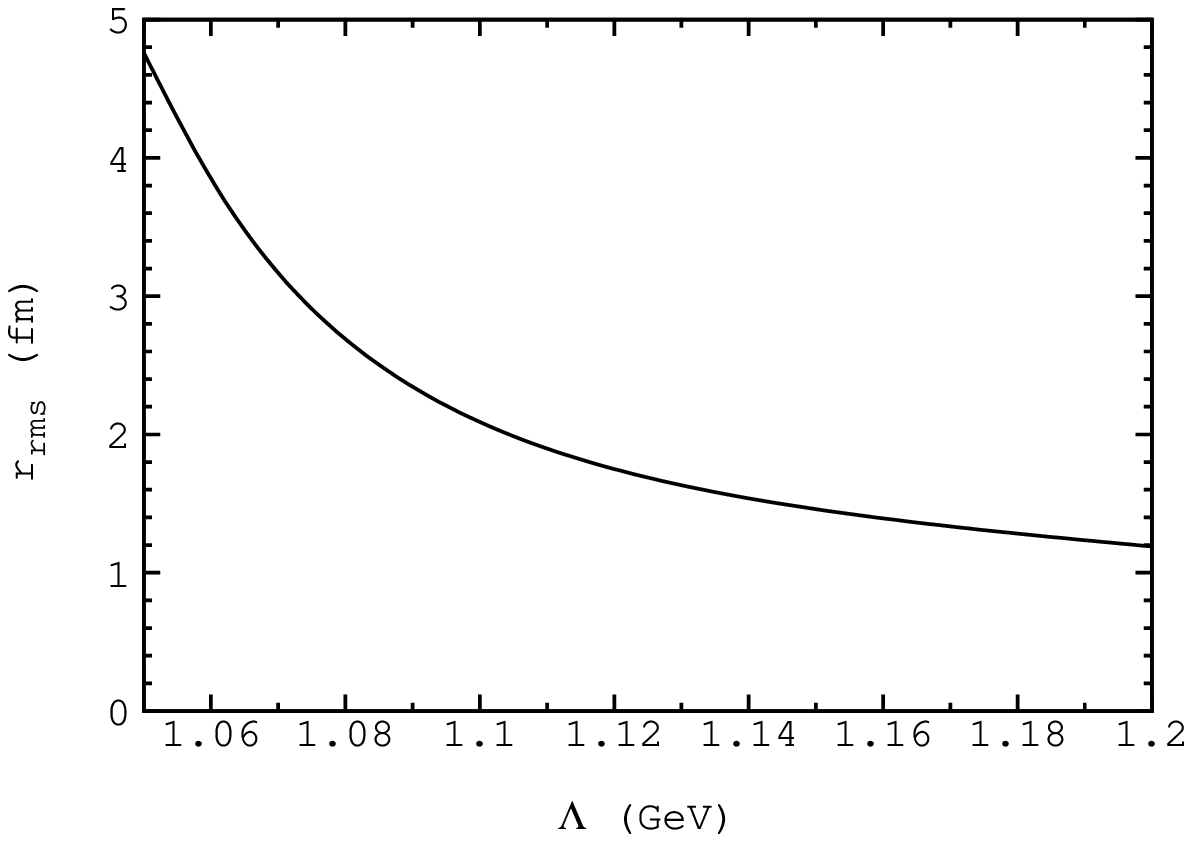}
\end{tabular}
\caption{The dependence of the binding solutions of $X(3872)$ with
$J^{PC}=1^{++}$ on the cutoff parameter. The left panel is for the
mass while the right panel is for the root-mean-square
radius.}\label{Plot:mass}
\end{figure}

If we further consider the isospin breaking, we obtain a loosely
bound state. When the cutoff parameter is fixed at 1.05 GeV, its
mass is 3871.51 MeV, which corresponds to the experimental value
of the mass of
$X(3872)$~\cite{Aaltonen:2009vj,Aaij:2011sn,Choi:2011fc}. The
root-mean-square radius is 4.76 fm which is larger than that of
the deuteron (about 2.0 fm). The dominant channel is ${1\over
\sqrt{2}}\left[D^{0}\bar{D}^{*0}-D^{*0}\bar{D}^{0}\right]|^3S_1>$,
with a probability of $86.80\%$. The second dominant channel is
${1\over \sqrt{2}}\left[D^{+}D^{*-}-D^{*+}D^{-}\right]|^3S_1>$, the
probability of which is $11.77\%$. And, the total probabilities of
the other channels is less than $1.5\%$. We plot the radial wave
functions of the individual channels in Fig.~\ref{Plot:Wave}. When
we increase the cutoff parameter to $1.10$ GeV, the mass of the
bound state decreases to 3869.28 MeV, and the root-mean-square
radius is 2.09 fm. The probability of the dominant channel
decreases to $70.44\%$ while that of the second dominant one increases
to $26.46\%$. In order to make clear the dependence of the binding
solution on the cutoff, we plot the variations of the mass and the
root-mean-square radius with the cutoff in Fig.~\ref{Plot:mass}.

\section{Isospin breaking in the hidden-charm decays of X(3872)}\label{isospin}

We focus on the isospin breaking in the wave function of
$X(3872)$. For simplicity, we analyze the numerical results in the
OBE model for illustration. Again, we adopt the following
short-hand notations, $\left[D^0\bar{D}^{*0}\right]\equiv{1\over
\sqrt{2}}\left(D^0\bar{D}^{*0}-D^{*0}\bar{D}^0\right)$,
$\left[D^+D^{*-}\right]\equiv{1\over
\sqrt{2}}\left(D^+D^{*-}-D^{*+}D^-\right)$,
$\left\{D^*\bar{D}^*\right\}\equiv{1\over
\sqrt{2}}\left(D^{*0}\bar{D}^{*0}+D^{*+}D^{*-}\right)$. The flavor
wave function of the $I=1, I_z=0$ state is $|10>={1\over
\sqrt{2}}\left(\left[D^+D^{*-}\right]-\left[D^0\bar{D}^{*0}\right]\right)$
while that of the isoscalar state is $|00>={1\over
\sqrt{2}}\left(\left[D^0\bar{D}^{*0}\right]+\left[D^+D^{*-}\right]\right)$.

The flavor wave function of the $X(3872)$ can be expanded as
\begin{eqnarray}
  X(3872)&=&{\chi_1(r)\over r}\left[D^0\bar{D}^{*0}\right]|^3S_1>
  +{\chi_2(r)\over r}\left[D^0\bar{D}^{*0}\right]|^3D_1>+{\chi_3(r)\over r}\left[D^+D^{*-}\right]|^3S_1> \nonumber\\
         &~&+{\chi_4(r)\over r}\left[D^+D^{*-}\right]|^3D_1>
         +{\chi_5(r)\over r}\left\{D^*\bar{D}^*\right\}|^3S_1>
         +{\chi_6(r)\over r}\left\{D^*\bar{D}^*\right\}|^3D_1> \nonumber \\
         &=&{1\over \sqrt{2}}{\chi_1(r)+\chi_3(r) \over r}|00>_{D\bar{D}^*}|^3S_1>
         +{1\over \sqrt{2}}{\chi_3(r)-\chi_1(r) \over r}|10>_{D\bar{D}^*}|^3S_1>
         +{1\over \sqrt{2}}{\chi_2(r)+\chi_4(r) \over r}|00>_{D\bar{D}^*}|^3D_1> \nonumber \\
         &~&+{1\over \sqrt{2}}{\chi_4(r)-\chi_2(r) \over r}|10>_{D\bar{D}^*}|^3D_1>
         +{\chi_5(r)\over r}|00>_{D^*\bar{D}^*}|^3S_1>
         +{\chi_6(r)\over r}|00>_{D^*\bar{D}^*}|^3D_1>.
\end{eqnarray}
So the probability of finding the isoscalar component within
$X(3872)$ is
\begin{eqnarray}
\rho_{00}&=&\int{\left[\chi_1(r)+\chi_3(r)\right]^2\over
2}dr+\int{\left[\chi_2(r)+\chi_4(r)\right]^2\over 2}dr
+\int\chi_5^2(r)dr+\int\chi_6^2(r)dr \nonumber \\
&=&\int\left[{\chi_1^2(r)+\chi_2^2(r)+\chi_3^2(r)+\chi_4^2(r)
\over 2}+\chi_1(r)\chi_3(r)+\chi_2(r)\chi_4(r)
+\chi_5^2(r)+\chi_6^2(r)\right]dr,
\end{eqnarray}
and the probability of finding the isovector component is
\begin{eqnarray}
\rho_{10}&=&\int{\left[\chi_3(r)-\chi_1(r)\right]^2\over
2}dr+\int{\left[\chi_4(r)-\chi_2(r)\right]^2\over 2}dr
\nonumber \\
&=&\int\left[{\chi_1^2(r)+\chi_2^2(r)+\chi_3^2(r)+\chi_4^2(r)
\over 2}-\chi_1(r)\chi_3(r)-\chi_2(r)\chi_4(r)\right]dr.
\end{eqnarray}

Numerically, the probability of the isoscalar component is
$73.76\%$ while that of the isovector component is $26.24\%$ if
the cutoff parameter is fixed at 1.05 GeV, which corresponds to a
tiny binding energy 0.3 MeV. However, if the binding energy
increases to $10.83$ MeV, the contribution of the isoscalar
component is as large as $98.51\%$ while that of the isovector
component is only $1.49\%$. In short, the isospin breaking depends
sensitively on the binding energy. To a large extent, the large
isospin symmetry breaking effect within the flavor wave functions
of $X(3872)$ is amplified by its tiny binding energy.

There exists strong experimental evidence that the decay of
$X(3872)\to J/\psi\pi^+\pi^-$ occurs through a virtual $\rho^0$
meson while the decay of $X(3872)\to J/\psi\pi^+\pi^-\pi^0$ occurs
through a virtual $\omega$ meson. We assume the decay $X(3872)\to
J/\psi\pi^+\pi^-$ comes from the $I=1$ component within the flavor
wave function of $X(3872)$ while $X(3872)\to J/\psi\pi^+\pi^-\pi^0$
comes from the $I=0$ component. In the present case, the different
phase space of the $J/\psi \rho$ and $J/\psi \omega$ decay modes
also plays an important role. Since the phase space is small, we
can safely neglect the higher partial waves and focus on the
S-wave decay only.  Now the ratio of these two phase space reads
\begin{eqnarray}
R_{Phase}={\int_{m_{3\pi}}^{m_{X(3872)}-m_{J/\psi}}dm_{\omega}\varrho(m_{\omega})|{\bf p}_{\omega}|\over
 \int_{m_{2\pi}}^{m_{X(3872)}-m_{J/\psi}}dm_{\rho}\varrho(m_{\rho})|{\bf p}_{\rho}|},
\end{eqnarray}
with
\begin{equation}
\varrho(m)={\Gamma\over 2\pi\left[(m-m_{cen})^2+{\Gamma^2 \over 4}\right]}
\end{equation}
being the mass distribution of the unstable particle and
\begin{equation}
|{\bf p}|={\left[\left(M^2_{X(3872)}-(m_{J/\psi}+m)^2\right)\left(M_{X(3872)}^2
-(m_{J/\psi}-m)^2\right)\right]^{1/2}\over 2M_{X(3872)}}
\end{equation}
being the decay momentum of the two-body decay.
The ratio of the isoscalar and isovector component within the
flavor wave functions of $X(3872)$ is defined as
\begin{equation}
R_I=\rho(I=0)/\rho(I=1).
\end{equation}
Finally we obtain the branching fraction ratio
\begin{equation}
R=R_{Phase}\times R_I = \mathcal{B}(X(3872)\to
\pi^+\pi^-\pi^0J/\psi)/\mathcal{B}(X(3872)\to \pi^+\pi^-
J/\psi)=0.42
\end{equation}
with the binding energy being 0.3 MeV.

Again, this ratio depends very sensitively on the binding energy
since the isospin breaking effect is very sensitive to the binding
energy. We provide several groups of the values of $R$ when the
binding energy varies from $0.1$ MeV to $1.0$ MeV in
Table~\ref{Branch}.

Given the uncertainty of experimental value of the mass of
$X(3872)$, this ratio is consisitent with the experimental value,
$1.0\pm0.4(\mbox{stat})\pm 0.3(\mbox{syst})$ from Belle Collaboration
~\cite{Abe:2005ix} and $0.8\pm 0.3$ from BABAR Collaboration
~\cite{delAmoSanchez:2010jr}.

\begin{table}[htp]
\caption{The binding solutions of $X(3872)$ with the OBE potential.
$\Lambda$ is the cutoff parameter. ``B.E." is the binding energy
while ``Mass" is the calculated mass of $X(3872)$. $r_{rms}$ and
``$P_i$" are the root-mean-square radius and the probability of
the ith channel, respectively. ``$\times$" means no binding
solutions, and ``$-$" denotes that the corresponding component
does not exist.}\label{Numerical:OBE}
  \begin{tabular*}{18cm}{@{\extracolsep{\fill}}ccccccccccc}
  \toprule[0.8pt]
  \toprule[0.8pt]\addlinespace[3pt]
    Cases       &$\L$ (GeV)& B.E. (MeV)& Mass (MeV)&$r_{rms}$ (fm)&$P_1(\%)$&
    $P_2(\%)$   & $P_3(\%)$& $P_4(\%)$ & $P_5(\%)$ & $P_6(\%)$   \\
    \specialrule{0.6pt}{3pt}{3pt}
    \multirow{4}*{I}&1.85  & $0.21$    & 3871.60   &   5.36       &99.54    &
        0.46    & $-$      & $ - $     &  $-$      &  $-$        \\
        ~       &    1.90  & $0.53$    & 3871.28   &   4.32       &99.27    &
        0.63    & $-$      & $ -$      &  $-$      &  $-$        \\
        ~       &    1.95  & $0.96$    & 3870.85   &   3.48       &99.18    &
        0.82    & $-$      & $ -$      &  $-$      & $-$         \\
        ~       &    2.00  & $1.51$    & 3870.30   &   2.88       & 98.99   &
        1.01    & $ -$     & $ -$      &  $-$      &  $-$        \\ [5pt]
\multirow{5}*{II}&    1.10 & $0.61$    & 3871.20   &   4.21
&98.82    &
          1.18  & $ -$     & $ -$      &  $-$      &  $-$        \\
        ~       &    1.15  & $2.15$    & 3869.66   &   2.54       &98.27    &
          1.73  & $ -$     & $-$       &  $-$      &  $-$        \\
        ~       &    1.20  & $4.58$    & 3867.23   &   1.84       &97.28   &
          2.18  & $ -$     & $ -$      &  $-$      &  $-$        \\
        ~       &    1.25  & $7.84$    & 3863.97   &   1.48       &97.40   &
          2.60  & $ -$     & $ -$      &  $-$      &  $-$        \\
        ~       &    1.30  & $11.87$   & 3859.94   &   1.26       &97.01   &
          2.99  & $ -$     & $ -$      &  $-$      &  $-$        \\  [5pt]
\multirow{6}*{III}&  1.00  & $0.74$    & 3871.07   &   3.92
&98.38   &
          0.79  &  $-$     &  $-$      &   0.66    &  0.18      \\
         ~      &    1.10  & $5.69$    & 3866.12   &   1.66       &96.39   &
          1.07  &   $-$    &  $-$      &    1.91   &  0.62       \\
        ~       &    1.15  & $9.67$    & 3862.14   &   1.34       &95.51   &
          1.12  &   $-$    &  $-$      &   2.46    &   0.92      \\
        ~       &    1.20  & $14.51$   & 3857.30   &   1.15       &94.65   &
          1.15  &   $-$    &  $-$      &  2.94     &   1.26      \\
        ~       &    1.25  & $20.18$   & 3851.63   &   1.02       &93.82   &
          1.17  &   $-$    &  $-$      &  3.35     &   1.67     \\
        ~       &    1.30  & $26.68$   & 3845.13   &   0.92       &92.98   &
          1.18  &   $-$    &  $-$      &   3.71    &  2.14       \\ [5.0pt]
\multirow{6}*{IV(Phy)}&   1.05  &  $0.30$   &  3871.51  &   4.76  &  86.80 &
 0.27           &  11.77  &   0.28    &   0.67    &  0.20      \\
       ~        &   1.06   &  $0.60$   &  3871.21  &   3.85       &  82.83 &
  0.33          &  15.35   &  0.34     &   0.88    &  0.27      \\
       ~        &  1.08    &  $1.43$   &  3870.38  &   2.69       &  75.80 &
  0.41          & 21.68    &  0.42     &   1.28    &  0.41      \\
      ~         &  1.10    &  $2.53$   &  3869.28  &   2.09       &  70.44 &
  0.46          & 26.46    &  0.47     &   1.62    &  0.54      \\
      ~         & 1.12     &  $3.84$   &  3867.97  &   1.75       &  66.40 &
  0.50          & 30.00    &  0.51     &   1.92    &  0.67      \\
      ~         & 1.15     &  $6.16$   &  3865.65  &   1.46       &  62.03 &
  0.53          & 33.72    &  0.54     &   2.31    &  0.87      \\
     ~          & 1.20     &  $10.83$  &  3860.98  &   1.19       &  57.38 &
  0.56          & 37.42    &  0.56     &   2.85    &  1.23      \\ [3pt]
 \bottomrule[0.8pt]
 \bottomrule[0.8pt]
\end{tabular*}
\end{table}

\begin{table}[htp]
\renewcommand{\arraystretch}{1.2}
\caption{The variation of the branching fraction ratio,
$R=\mathcal{B}(X(3872)\to
\pi^+\pi^-\pi^0J/\psi)/\mathcal{B}(X(3872)\to\pi^+\pi^-J/\psi)$,
with the binding energy. ``$R_{Phase}$" is the ratio of the phase
space between $J/\psi \omega$ and $J/\psi \rho$.
$R_I=\rho(I=0)/\rho(I=1)$ is the ratio of the isoscalar and
isovector component.}\label{Branch}
 \begin{tabular*}{18cm}{@{\extracolsep{\fill}}cccc}
  \toprule[0.8pt]
  \toprule[0.8pt]\addlinespace[3pt]
  B.E.(MeV)     &  $R_{Phase}$ &  $R_I$       &  $R$ \\
\specialrule{0.6pt}{3pt}{3pt}
 0.10           &   0.154      &  65.76/34.04 &  0.30     \\
 0.20           &   0.153      &  70.05/29.95 &  0.36     \\
 0.30           &   0.152      &  73.76/26.24 &  0.42     \\
 0.60           &   0.150      &  79.81/20.19 &  0.59     \\
 1.00           &   0.147      &  84.32/15.68 &  0.79     \\ [3pt]
\bottomrule[0.8pt] \bottomrule[0.8pt]
\end{tabular*}
\end{table}

\section{Conclusions}\label{Summary}

In the present work, we have performed an extensive study of the
possibility of $X(3872)$ as a $J^{PC}=1^{++}$ ``hadronic molecule"
with both the OPE and OBE potential. We have considered the
contribution from the light meson exchange including $\pi$,
$\eta$, $\sigma$, $\rho$ and $\omega$. It is important to note that
the contribution from the heavier $\eta$, $\sigma$, $\rho$ and
$\omega$ exchanges cancels each other to a very large extent. As a
consequence, the long-range pion exchange plays a dominant role in
forming the loosely bound $X(3872)$ state.

We have considered the S-D wave mixing which plays an important
role in the deuteron case, the charged $D\bar{D}^*$ mode, the
coupling of $D\bar{D}^*$ to $D^*\bar{D}^*$ and the isospin
breaking coming from the mass difference between the neutral and
charged $D(D^*)$ meson. All these factors play an important role
in forming the loosely bound $X(3872)$.

The inclusion of the charged $D\bar{D}^*$ mode enhances the
attraction. Now there exists one loosely bound isoscalar state
with a reasonable cutoff around 1.1 GeV within the OBE model. The
coupling of $D\bar{D}^*$ to $D^*\bar{D}^*$ will further enhance
the binding force and increase the binding energy by around 5 MeV
with the same parameter. However the resulting state is still an
isoscalar. If we take into account the 8 MeV mass difference
between the charged and neutral $D\bar{D}^*$ pairs, the binding energy
decreases by about 3 MeV. Our numerical analysis indicates that
the hadronic molecule with $J^{PC}=1^{++}$ in Case IV can be
identified as the physical $X(3872)$ state.

If we take the binding energy of $X(3872)$ as 0.3 MeV, the channel
${1\over
\sqrt{2}}\left[D^{0}\bar{D}^{*0}-D^{*0}\bar{D}^{0}\right]|^3S_1>$ is
dominant with a largest probability $86.80\%$, while that of the
channel ${1\over \sqrt{2}}\left[D^{+}D^{*-}-D^{*+}D^{-}\right]|^3S_1>$
is $11.77\%$. Moreover, our results indicate that there exists
large isospin breaking in the flavor wave function of $X(3872)$.
The isospin breaking depends strongly on the binding energy. The
deeper the binding is, the smaller the isospin breaking effect
becomes. When the binding energy is 0.30 MeV, the probabilities of
the isoscalar and isovector components are around $73.76\%$ and
$26.24\%$ respectively. However, they change to about $98.51\%$
and $1.49\%$ respectively when the binding energy increases to
10.83 MeV. The extreme sensitivity of the physical observables to
the binding energy is characteristic of the very loosely bound
system.

Taking into account the phase space difference as well as the
isospin breaking, we obtain the branching fraction ratio
$R=\mathcal{B}(X(3872)\to
\pi^+\pi^-\pi^0J/\psi)/\mathcal{B}(X(3872)\to \pi^+\pi^-
J/\psi)=0.42$ for the binding energy being around 0.3 MeV, which
is in rough agreement with the current experimental measurement
from Belle and Babar Collaborations.

Recently Faccini {\it et al.} have performed a study of the spin of
$X(3872)$ by fitting the experimental data. In their combined fit,
they excluded the $2^{-+}$ hypothesis at $99.9\%$ C.L., but
returns a probability of only $5.5\%$ of the $1^{++}$ hypothesis
being correct. However, in their separate fit they obtained a
preference for the $1^{++}$ hypothesis in the $J/\psi \rho$
channel with a probability of $23\%$ and an $81\%$ preference for
the $2^{-+}$ assignment in the $J/\psi \omega$
channel~\cite{Faccini:2012zv}.

We have also studied the possibility of the $2^{-+}$ assignment of
the $X(3872)$. We considered the charged mode of $D\bar{D}^*$, the
isospin-breaking and the coupling to $D^*\bar{D}^*$. But, we find
no binding solutions with a reasonable cutoff parameter less than
2.0 GeV. In the $1^{++}$ case, the coupling to $D^*\bar{D}^*$
increases the binding energy by a few MeV. One may also wonder
about the possibility of $X(3872)$ being a deeply bound P-wave
$D^*\bar{D^*}$ state. Therefore we have also investigated the
$D^*\bar{D}^*$ channel with explicit isospin-breaking and the
$P$-wave and $F$-wave mixing. We obtained a deeply bound state
with mass $3871.61$ MeV when we tuned the cutoff parameter to be
as high as 2.642 GeV. With so large a binding energy (142.33 MeV),
the isospin breaking effect almost disappears completely. Its
flavor wave function is an isoscalar, which is in conflict with
the experimental result. It seems that the $2^{-+}$ assignment of $X(3872)$ is not
favored within the present meson exchange model. Certainly
the investigation of the $2^{-+}$ possibility with other
theoretical approaches will be very helpful to settle this issue.

In short summary, the existence of the loosely bound X(3872) state
and the large isospin symmetry breaking in its hidden-charm decay
arises from the combined and very delicate efforts of the several
driving forces including the long-range one-pion exchange, the S-D
wave mixing, the mass splitting between the charged and neutral
$D(D^\ast)$ mesons, and the coupled-channel effects.

\section*{Acknowledgments}

This project was supported by the National Natural Science
Foundation of China under Grants 11075004, 11021092 and Ministry
of Science and Technology of China (2009CB825200). This work is
also supported in part by the DFG and the NSFC through funds
provided to the sino-germen CRC 110 ``Symmetries and the Emergence
of Structure in QCD''.


\begin{thebibliography}{99}

 \bibitem{Abe:2003hq}
K.~Abe  {\it et~al.}, Belle Collaboration,
 \newblock Phys.~Rev.~Lett. {\bf 91}, 262001 (2003).

 \bibitem{Acosta:2003zx}
D.~Acosta  {\it et~al.}, CDF Collaboration,
 \newblock Phys.~Rev.~Lett. {\bf 93}, 072001 (2004).

 \bibitem{Abazov:2004kp}
V.~Abazov  {\it et~al.}, D0 Collaboration,
 \newblock Phys.~Rev.~Lett. {\bf 93}, 162002 (2004).

 \bibitem{Aubert:2004fc}
B.~Aubert  {\it et~al.}, BABAR Collaboration,
 \newblock Phys.~Rev.~Lett. {\bf 93}, 041801 (2004).

 \bibitem{Aaij:2011sn}
R.~Aaij  {\it et~al.}, LHCb Collaboration,
 \newblock (2011), arXiv:1112.5310 [hep-ex].

 \bibitem{Choi:2011fc}
S.-K. Choi  {\it et~al.},
 \newblock Phys.~Rev. {\bf D84}, 052004 (2011).

 \bibitem{Braaten:2004jg}
E.~Braaten,
 \newblock Phys.Rev. {\bf D73}, 011501 (2006).

 \bibitem{Matheus:2006xi}
R.~D. Matheus, S.~Narison, M.~Nielsen, and J.~Richard,
 \newblock Phys.Rev. {\bf D75}, 014005 (2007).

 \bibitem{Chiu:2006hd}
T.-W. Chiu and T.-H. Hsieh, TWQCD Collaboration,
 \newblock Phys.Lett. {\bf B646}, 95 (2007).

 \bibitem{Liu:2008qb}
Y.-R. Liu and Z.-Y. Zhang,
 \newblock Phys.Rev. {\bf C79}, 035206 (2009).

 \bibitem{Gamermann:2009fv}
D.~Gamermann and E.~Oset,
 \newblock Phys.Rev. {\bf D80}, 014003 (2009).


 \bibitem{Matheus:2009vq}
R.~D. Matheus, F.~Navarra, M.~Nielsen, and C.~Zanetti,
 \newblock Phys.Rev. {\bf D80}, 056002 (2009).

 \bibitem{DeSanctis:2011zza}
M.~De~Sanctis and P.~Quintero,
 \newblock Eur.Phys.J. {\bf A47}, 54 (2011).

 \bibitem{Valderrama:2012jv}
M.~P. Valderrama,
 \newblock Phys.Rev. {\bf D85}, 114037 (2012).

 \bibitem{Yang:2012my}
Y.-B. Yang  {\it et~al.},
 \newblock (2012), arXiv:1206.2086 [hep-lat].


 \bibitem{Swanson:2003tb}
E.~S. Swanson,
 \newblock Phys.~Lett. {\bf B588}, 189 (2004).

 \bibitem{Tornqvist:2004qy}
N.~A. T\"ornqvist,
 \newblock Phys.~Lett. {\bf B590}, 209 (2004).

 \bibitem{AlFiky:2005jd}
M.~T. AlFiky, F.~Gabbiani, and A.~A. Petrov,
 \newblock Phys.~Lett. {\bf B640}, 238 (2006).

 \bibitem{Thomas:2008ja}
C.~E. Thomas and F.~E. Close,
 \newblock Phys.~Rev. {\bf D78}, 034007 (2008).

 \bibitem{Liu:2008tn}
X.~Liu, Z.-G. Luo, Y.-R. Liu, and S.-L. Zhu,
 \newblock Eur.~Phys.~J. {\bf C61}, 411 (2009).

 \bibitem{Lee:2009hy}
I.~W. Lee, A.~Faessler, T.~Gutsche, and V.~E. Lyubovitskij,
 \newblock Phys.~Rev. {\bf D80}, 094005 (2009).

 \bibitem{Li:2004sta}
B.~A. Li,
 \newblock Phys.~Lett. {\bf B605}, 306 (2005).

 \bibitem{Maiani:2004vq}
L.~Maiani, F.~Piccinini, A.~D.~Polosa, and V.~Riquer,
 \newblock Phys.~Rev. {\bf D71}, 014028 (2005).

 \bibitem{Barnes:2003vb}
T.~Barnes and S.~Godfrey,
 \newblock Phys.~Rev. {\bf D69}, 054008 (2004).

 \bibitem{Suzuki:2005ha}
M.~Suzuki,
 \newblock Phys.~Rev. {\bf D72}, 114013 (2005).

 \bibitem{Aubert:2005vi}
B.~Aubert  {\it et~al.}, BABAR Collaboration,
 \newblock Phys.~Rev.~Lett. {\bf 96}, 052002 (2006).

 \bibitem{Abe:2005ix}
K.~Abe  {\it et~al.}, Belle Collaboration,
 \newblock (2005), arXiv:hep-ex/0505037 [hep-ex].

 \bibitem{delAmoSanchez:2010jr}
P.~del Amo~Sanchez  {\it et~al.}, BABAR Collaboration,
 \newblock Phys.~Rev. {\bf D82}, 011101 (2010).

 \bibitem{Aubert:2006aj}
B.~Aubert  {\it et~al.}, BABAR Collaboration,
 \newblock Phys.~Rev. {\bf D74}, 071101 (2006).

 \bibitem{Abe:2005iya}
K.~Abe  {\it et~al.}, Belle Collaboration,
 \newblock (2005), arXiv:hep-ex/0505038 [hep-ex].

 \bibitem{Abulencia:2006ma}
A.~Abulencia  {\it et~al.}, CDF Collaboration,
 \newblock Phys.~Rev.~Lett. {\bf 98}, 132002 (2007).

 \bibitem{Falk:1992cx}
A.~F. Falk and M.~E. Luke,
 \newblock Phys.~Lett. {\bf B292}, 119 (1992).

 \bibitem{PhysRevD.45.R2188}
M.~B. Wise,
 \newblock Phys.~Rev. {\bf D45}, R2188 (1992).

 \bibitem{Yan1992}
T.-M. Yan  {\it et~al.},
 \newblock Phys.~Rev. {\bf D46}, 1148 (1992).

 \bibitem{Grinstein:1992qt}
B.~Grinstein, E.~E. Jenkins, A.~V. Manohar, M.~J. Savage, and M.~B. Wise,
 \newblock Nucl.~Phys. {\bf B380}, 369 (1992).

 \bibitem{Cheng:1992xi}
H.-Y. Cheng  {\it et~al.},
 \newblock Phys.~Rev. {\bf D47}, 1030 (1993).

 \bibitem{Casalbuoni:1996pg}
R.~Casalbuoni  {\it et~al.},
 \newblock Phys.~Rept. {\bf 281}, 145 (1997).

 \bibitem{Sun2011}
Z.-F. Sun, J.~He, X.~Liu, Z.-G. Luo, and S.-L. Zhu,
 \newblock Phys.~Rev. {\bf D84}, 054002 (2011).

 \bibitem{Ding2009a}
G.-J. Ding,
 \newblock Phys.~Rev. {\bf D79}, 014001 (2009).

 \bibitem{Falk:1992cx}
A.~F. Falk and M.~E. Luke,
 \newblock Phys.Lett. {\bf B292}, 119 (1992).

 \bibitem{Dai:1998vh}
Y.-B. Dai and S.-L. Zhu,
 \newblock Eur.Phys.J. {\bf C6}, 307 (1999).

 \bibitem{Navarra:2000ji}
F.~Navarra, M.~Nielsen, M.~Bracco, M.~Chiapparini, and C.~Schat,
 \newblock Phys.Lett. {\bf B489}, 319 (2000).

 \bibitem{Ahmed:2001xc}
S.~Ahmed  {\it et~al.}, CLEO Collaboration,
 \newblock Phys.~Rev.~Lett. {\bf 87}, 251801 (2001).



 \bibitem{Isola:2003fh}
C.~Isola, M.~Ladisa, G.~Nardulli, and P.~Santorelli,
 \newblock Phys.~Rev. {\bf D68}, 114001 (2003).

\bibitem{Bando:1987br}
  M.~Bando, T.~Kugo and K.~Yamawaki,
   Phys.\ Rept.\  {\bf 164}, 217 (1988).

 \bibitem{Liu2008a}
X.~Liu, Y.-R. Liu, W.-Z. Deng, and S.-L. Zhu,
 \newblock Phys.~Rev. {\bf D77}, 094015 (2008).

\bibitem{pdg2010}
K.~Nakamura, {\em et al.} (Particle Data Group), J.~Phys. {\bf G37}, 075021 (2010).

 \bibitem{Liu:2008fh}
Y.-R. Liu, X.~Liu, W.-Z. Deng, and S.-L. Zhu,
 \newblock Eur.~Phys.~J. {\bf C56}, 63 (2008).

 \bibitem{Abrashkevich199565}
A.~Abrashkevich, D.~Abrashkevich, M.~Kaschiev, and I.~Puzynin,
 \newblock Computer Physics Communications {\bf 85}, 65  (1995).

 \bibitem{Abrashkevich199890}
A.~Abrashkevich, D.~Abrashkevich, M.~Kaschiev, and I.~Puzynin,
 \newblock Computer Physics Communications {\bf 115}, 90  (1998).

\bibitem{Aceti:2012cb}
  F.~Aceti, R.~Molina and E.~Oset,
  arXiv:1207.2832 [hep-ph].

\bibitem{Artoisenet:2010va}
  P.~Artoisenet, E.~Braaten and D.~Kang,
  Phys.\ Rev.\ {\bf D82}, 014013 (2010).

\bibitem{Braaten:2007ft}
  E.~Braaten and M.~Lu,
   Phys.\ Rev.\  {\bf D77}, 014029 (2008).


 \bibitem{Aaltonen:2009vj}
T.~Aaltonen  {\it et~al.}, CDF Collaboration,
 \newblock Phys.~Rev.~Lett. {\bf 103}, 152001 (2009).

\bibitem{Faccini:2012zv}
  R.~Faccini, F.~Piccinini, A.~Pilloni and A.~D.~Polosa,
  Phys.\ Rev.\  {\bf D86}, 054012 (2012).



\end{thebibliography}


\section{APPENDIX}\label{Appendix}

\subsection{Some Helpful Functions}\label{Function}

The functions $H_i$ etc are defined as,
\begin{eqnarray}
H_0(\Lambda,m,r)&=&Y(u r)-\frac{\chi}{u}Y(\chi r)
-\f{r\beta^2}{2u}Y(\chi r), \qquad H_1(\L,m,r)=Y(u
r)-\f{\chi}{u}Y(\chi r)
-\f{r\chi^2\beta^2}{2u^3}Y(\chi r), \n\\
H_3(\L,m,r)&=&Z(u r)-\f{\chi^3}{u^3}Z(\chi r) -\f{\chi
\beta^2}{2u^3}Z_2(\chi r),\qquad M_1(\L,m,r)=-\f{1}{\theta
r}\left[\cos(\theta r)-e^{-\chi r}\right]
-\f{\chi\beta^2}{2\theta^3}e^{-\chi r},\n\\
M_3(\L,m,r)&=&-\left[\cos{(\theta r)}-\f{3\sin{(\theta r)}}{\theta
r} -\f{3\cos{(\theta r)}}{\theta^2r^2}\right]\f{1}{\theta r}
-\f{\chi^3}{\theta^3}Z(\chi r)-\f{\chi\beta^2}{2\theta^3}Z_2(\chi
r),
\end{eqnarray}
where,
\begin{eqnarray*}
 \beta^2=\L^2-m^2,\quad u^2=m^2-Q_0^2,\quad \theta^2=-(m^2-q_0^2),
\quad\chi^2=\L^2-q_0^2,
\end{eqnarray*}
and
\begin{eqnarray*}
 Y(x)=\f{e^{-x}}{x},\quad Z(x)=\left(1+\f{3}{x}+\f{3}{x^2}\right)Y(x),\quad
 Z_1(x)=\left(\f{1}{x}+\f{1}{x^2}\right)Y(x),\quad Z_2(x)=(1+x)Y(x).
\end{eqnarray*}
Fourier transformation formulae read:
\begin{eqnarray}
\f{1}{u^2+\bm{q}^2}&\rightarrow&\f{u}{4\pi}H_0(\L,m,r),\quad
\f{\bm{q}^2}{u^2+\bm{q}^2}\rightarrow-\f{u^3}{4\pi}H_1(\L,m,r), \n\\
\f{\bm{q}}{u^2+\bm{q}^2}&\rightarrow&\f{iu^3}{4\pi}\bm{r}H_2(\L,m,r),\quad
\f{q_iq_j}{u^2+\bm{q}^2}\rightarrow-\f{u^3}{12\pi}\left[H_3(\L,m,r)
k_{ij}+H_1(\L,m,r)\delta_{ij}\right], \label{FTformula}
\end{eqnarray}
where, $k_{ij}=3\f{r_ir_j}{r^2}-\delta_{ij}.$

\subsection{The Possible $D\bar{D}^*$ molecular State With $J^{PC}=1^{-+}$}\label{1^-+}

As a byproduct, we extend our formalism to the $J^{PC}=1^{-+}$
case and collect the numerical results in
Table~\ref{numerical:1mp}. The flavor wave function of the state
with $J^{PC}=1^{-+}$ (denoted as $X^{'}$) is
\begin{eqnarray}
  X^{'}=\chi_1(r){1\over \sqrt{2}}\left(D^{*0}\bar{D}^0-D^0\bar{D}^{*0}\right)|^3P_1>
  +\chi_2(r){1\over \sqrt{2}}\left(D^{*+}D^--D^+D^{*-}\right)|^3P_1>
  +\chi_3(r){1\over \sqrt{2}}\left(D^{*0}\bar{D}^{*0}+D^{*+}D^{*-}\right)|^3P_1>
\end{eqnarray}
The other three channels, ${1\over
\sqrt{2}}\left(D^{*0}\bar{D}^{*0}+D^{*+}D^{*-}\right)|^1P_1>$,
${1\over
\sqrt{2}}\left(D^{*0}\bar{D}^{*0}+D^{*+}D^{*-}\right)|^5P_1>$ and
${1\over
\sqrt{2}}\left(D^{*0}\bar{D}^{*0}+D^{*+}D^{*-}\right)|^5F_1>$ have
been omitted with the same reason as for the $J^{PC}=1^{++}$ case.

We obtain a loosely bound state with binding energy 1.60
MeV and root-mean-square radius 1.49 fm when the cutoff parameter
is fixed at 1.80 GeV,  The probabilities of the channels
${1\over \sqrt{2}}\left(D^{*0}\bar{D}^0-D^0\bar{D}^{*0}\right)|^3P_1>$ and
${1\over \sqrt{2}}\left(D^{*+}D^--D^+D^{*-}\right)|^3P_1>$ are
$56.66\%$ and $41.06\%$, respectively~\ref{numerical:1mp}.
However, when we increase the cutoff parameter to 1.90 GeV, the
binding energy increases sharply to 23.66 MeV, and the
root-mean-square radius decreases to 0.83 fm, which suggests that
the results depend sensitively on the cutoff parameter.
\begin{table}[htp]
\renewcommand{\arraystretch}{1.2}
 \caption{The numerical results of the state with $J^{PC}=1^{-+}$ with the
OBE potential.}\label{numerical:1mp}
\begin{tabular*}{18cm}{@{\extracolsep{\fill}}ccccccc}
 \toprule[1.0pt]\addlinespace[3pt]
 $\Lambda$ (GeV)  &  B.E. (MeV) &   Mass & $r_{rms}$ (fm)& $P_1(\%)$ & $P_2(\%)$ &  $P_3(\%)$ \\
 \specialrule{0.6pt}{3pt}{3pt}
 1.80             &  1.60       &3870.21 & 1.49          &  56.66    & 41.06     &  2.28      \\
 1.82             &  5.04       &3866.77 & 1.17          &  54.05    & 43.26     &  2.69      \\
 1.84             &  8.98       &3862.83 & 1.04          &  52.66    & 44.28     &  3.06      \\
 1.86             & 13.41       &3858.40 & 0.95          &  51.71    & 44.87     &  3.42      \\
 1.88             & 18.30       &3853.51 & 0.88          &  50.98    & 45.24     &  3.78      \\
 1.90             & 23.66       &3848.15 & 0.83          &  50.39    & 45.46     &  4.14      \\ [3pt]
 \bottomrule[1.0pt]
\end{tabular*}
\end{table}

\end{document}